\documentclass[preprint2]{aastex62}
\usepackage{natbib}

\graphicspath{{./}{figures/}}

\begin{document}

\title{Asymmetric Drift in the Andromeda Galaxy (M31) as a Function of Stellar Age}

\correspondingauthor{Amanda C. N. Quirk}
\email{acquirk@ucsc.edu}

\author[0000-0001-8481-2660]{Amanda Quirk}
\affil{UCO/Lick Observatory, University of California at Santa Cruz \\
1156 High Street
Santa Cruz, CA 95064, USA}

\author{Puragra Guhathakurta}
\affiliation{UCO/Lick Observatory, University of California at Santa Cruz \\
1156 High Street
Santa Cruz, CA 95064, USA}

\author{Laurent Chemin}
\affiliation{Centro de Astronom\'ia, Universidad de Antofagasta \\
Avda. U. de Antofagasta 02800, Antofagasta, Chile}

\author{Claire E. Dorman}
\affiliation{UCO/Lick Observatory, University of California at Santa Cruz \\
1156 High Street
Santa Cruz, CA 95064, USA}

\author[0000-0003-0394-8377]{Karoline M. Gilbert}
\affiliation{Space Telescope Science Institute, 3700 San Martin Dr., Baltimore, MD 21218, USA}
\affiliation{Department of Physics \& Astronomy, Bloomberg Center for Physics and Astronomy, Johns Hopkins University, 3400 N. Charles Street, Baltimore, MD 21218}

\author{Anil C. Seth}
\affiliation{Department of Physics and Astronomy, University of Utah, \\ Salt Lake City, UT 84112, USA}

\author{Benjamin F. Williams}
\affiliation{Department of Astronomy, University of Washington, \\
Box 351580, Seattle, WA 98195, USA}

\author{Julianne J. Dalcanton}
\affiliation{Department of Astronomy, University of Washington, \\
Box 351580, Seattle, WA 98195, USA}

\begin{abstract}

We analyze the kinematics of Andromeda's disk as a function of stellar age by using photometry from the Panchromatic Hubble Andromeda Treasury (PHAT) survey and spectroscopy from the Spectroscopic and Photometric Landscape of Andromeda's Stellar Halo (SPLASH) survey. We use HI 21-cm and CO ($\rm J=1 \rightarrow 0$) data to examine the difference between the deprojected rotation velocity of the gas and that of the stars. We divide the stars into four stellar age bins, from shortest lived to longest lived: massive main sequence stars (0.03 Gyr), more luminous intermediate mass asymptotic giant branch (AGB) stars (0.4 Gyr), less luminous intermediate mass AGB stars (2 Gyr), and low mass red giant branch stars (4 Gyr). There is a clear correlation between the offset of the stellar and the gas rotation velocity, or the asymmetric drift: the longer lived populations lag farther behind the gas than short lived populations. We also examine possible causes of the substructure in the rotation curves and find that the most significant cause of scatter in the rotation curves comes from the tilted ring model being an imperfect way to account for the multiple warps in Andromeda's disk. 

\end{abstract}


\section{Introduction} \label{sec:intro}
\par The kinematics of stars are shaped by a galaxy's heating history. The properties of present day galaxies cannot be explained without dynamical heating events \citep{Seth,Walker1996}. Spiral waves in the disks of galaxies have long been suspected as a source of heating, but in practice cause few changes in the amount of random stellar motions in a galaxy \citep{Sellwood2002,Sellwood2013}. Instead, mergers of satellite galaxies are needed to explain the stellar vertical velocity dispersions seen in the spiral galaxies of the Local Group \citep{Leaman2017}, as the mergers can stir up the disk and perturb the orbits of stars \citep{Quinn1986}. 

\par The dynamical effects of these events remain long after the event itself and result in greater velocity dispersion and thicker disks.
Dynamical heating permanently disrupts the orbital paths of stars \citep{Leaman2017} because stars are collisionless and are heated. Dynamical heating events cause stars to be displaced to greater radii, and conservation of angular momentum dictates that their orbital velocity decreases during this migration: $L=mvr$ \citep{Binney2008}. 
\cite{Sellwood2002} find that changes in the angular momentum distribution always increase random motion in stellar populations, which results in stars having greater velocity dispersion than gas. Gas is collisional and therefore can maintain more uniform circular orbits \citep{Sellwood11999}. Thus, while stars have greater velocity dispersion than the surrounding gas, they tend to have lower circular velocities because the circular velocity is a smaller component of their total velocity. 

\par The difference between the gas rotation velocity and the stellar rotation velocity at a certain radii is called asymmetric drift ($v_{a}$, lag, or AD) \citep{Stromberg}. 
AD is a proxy for heating events because it suggests stars are on perturbed non-circular orbits. In addition to reflecting the accretion history of a galaxy, AD is also an important tool for learning about the current state of a galaxy. Specifically, it reflects the dynamics of the galactic disk and can be used to measure stability criterion \citep{Westfall2007}. Additionally, lag is a proxy for interactions with a bar \citep{Dehnen1998a} and for velocity dispersion. \cite{Dorman2015} shows velocity dispersion increases with stellar age, for stars that are longer lived have experienced more dynamical heating. Since lag is a proxy for velocity dispersion, we expect it to similarly increase with stellar age \citep{Walker1996, Dehnen1998a,Dehnen1998}. This lag will be preserved in older populations because these populations are dynamically relaxed. 

\par Making AD calculations requires measurements of a galaxy's stellar and gas motions. Integral-field-unit (IFU) spectroscopy allows astronomers to study the kinematics of galaxies by obtaining spectra along two dimensions \citep{Martinsson2013} and are powerful enough to make AD measurements for even distant galaxies \citep{Bershady2010}. \cite{Martinsson2013} use IFUs from SparsePak \citep{Bershady2004,Bershady2005} to measure the AD of stars compared to ionized gas (OIII) in face on spirals and find that stars lagged behind the gas on average by $11~\pm8~\%$. This value is similar to other studies of lag in local galaxies \citep{Ciardullo2004,Herrmann2009,Westfall2007, Westfall2011} and in the Milky Way (MW) \citep{Ratnatunga1997,Olling}. The Mapping Nearby Galaxies at Apache Point Observatory survey (MaNGA) will extend the sample of galaxies for which we have AD measurements \citep{Bundy2015}. IFU spectroscopy, however, cannot resolve individual stars in distant galaxies. In our own galaxy, on the other hand, we can measure the velocities of individual stars, which has allowed studies to account for AD when constructing rotation curves (RCs) of the MW \citep{Golubov2014,Huang2015}. Lag has also been used to determine the local standard of rest and to identify warps in the MW, which could possibly be explained by dynamical interactions with the Galactic bar \citep{Dehnen1998a}.  

\par Studies of AD in the MW are difficult and incomplete though, as we are limited to the Solar neighborhood \citep{Dehnen1998a}. Examining the Andromeda galaxy or Messier 31 (M31) can give us a more complete idea of AD in a spiral galaxy because at a distance of 785~Mpc \citep{Mcconnachie2005}, we can measure velocities of individual stars and thus the stellar rotation and velocity dispersion. The main objective of this article is to compare the rotation velocity of four populations of stars in M31 to that of the gas, which is assumed to trace more circular motions. This gives us the opportunity to measure AD as function of stellar age and to verify the extent to which AD is a proxy for velocity dispersion \citep{Bershady2010}. 

\par M31 is an interesting candidate for an AD study because it has experienced heating events throughout its lifetime. The outer halo of M31 shows tidal streams, which are believed to be relics of minor mergers \citep{Hernquist1988,Hernquist1989}. The mass of the stellar halo, on the other hand, can more easily be explained by a major merger \citep{Bell2017, DSouza2018, Hammer2018}. A major merger could also trigger the burst of star formation that occurred $\sim$2~Gyr ago \citep{Williams2017, Williams2018, DSouza2018}. The most recent major event, which created the Giant Stellar Stream (GSS), is believed to have occurred $\sim$1~Gyr ago and enriched the galaxy's inner halo \citep{Ibata2001, Ibata, Ferguson}. These accretion events have affected the dynamics of M31 and have left it with a higher local velocity dispersion and a wider distribution of dispersion across the disk than that in the MW \citep{Sysoliatina2018,Budanova2017}.
This suggests M31 has had a more violent history or a prolonged period of accretion than the MW \citep{Ferguson2016}.

\par This paper is organized as follows: in Section~\ref{sec:Data}, we describe the four datasets we use in this analysis. In Section~\ref{sec:AD_age}, we explain how our stellar sample is divided into four age groups and the analysis of the dynamics and AD of each age bin. In Section~\ref{sec:RC_messy}, we analyze possible causes for additional structure seen in the RCs. The sources of substructure we examine include geometrical effects from the tilted ring model (Section~\ref{sec:geometry}) and the multiplicity in the HI (Section~\ref{sec:HI}). Other possible sources are included in the Appendix.
In Section~\ref{sec:Summary}, we summarize our results. 

\section{Data}\label{sec:Data}
\par
\begin{figure}[h!] 
\epsscale{1.25}
\plotone{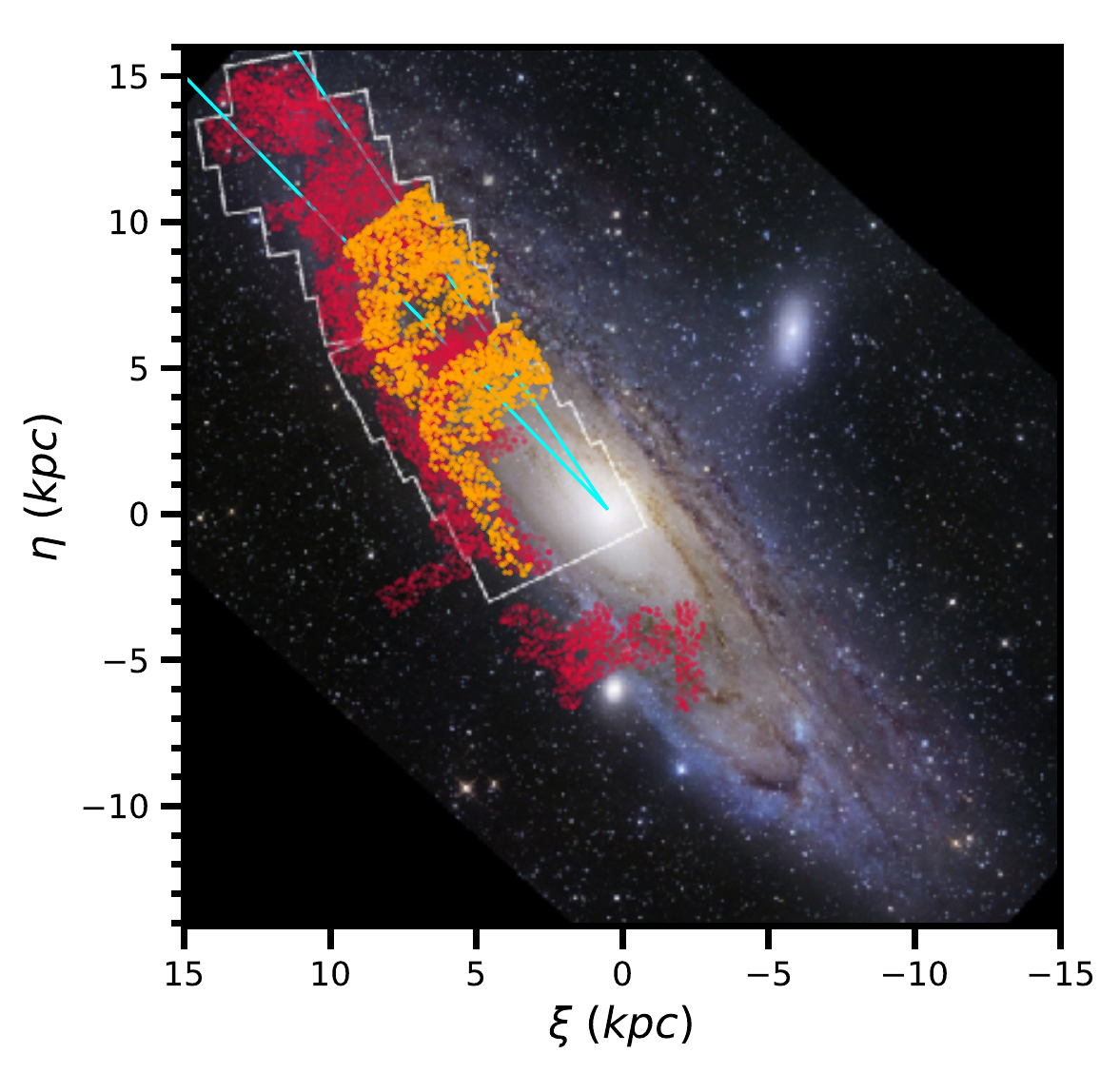}
\caption{Map of survey coverage for the four datasets. The white outline represents the PHAT survey, the orange the CO coverage, and the red points show the SPLASH survey area. The HI survey covers the area shown in the image and extends beyond it. The aqua lines denote the wedge in which the tilted ring deprojection factor is small. This is further discussed in Section~\ref{sec:geometry}. The image of M31 was taken by Robert Gendler. \label{fig:surveys}}
\end{figure}

 This study uses four datasets from various surveys of M31: (1)~spectroscopy from the Keck~II telescope and DEep Imaging Multi-Object Spectrograph (DEIMOS) obtained as part of the Spectroscopic and Photometric Landscape of Andromeda's Stellar Halo  (SPLASH) survey \citep{Guhathakurta2005,Guhathakurta2006,Dorman2015}, (2)~photometry from Hubble Space Telescope (HST) and Advanced Camera for Surveys (ACS) and Wide Field Camera~3 (WFC3) imagers as part of the Panchromatic Hubble Andromeda Treasury (PHAT) survey \citep{Dalcanton2012, Williams2014}, (3)~HI 21-cm data obtained at the Dominion Radio Astrophysical Observatory \citep{Chemin2006}, and (4)~CO $\rm J=1 \rightarrow 0$ emission obtained by the IRAM telescope \citep{Nieten2006}. 
The HI data covers the full area of the HI disk, which includes all of the bright stellar disk. Of the four surveys, the CO data covers the smallest area, and the spatial coverage of the SPLASH survey in M31's disk is somewhat larger than that of the PHAT survey. Figure~\ref{fig:surveys} compares the area coverage of the PHAT and SPLASH surveys to the CO coverage. Each dataset is briefly described in the following subsections. For more details about the target selection, observations, and data collection, see the papers cited for each survey.

\subsection{Keck DEIMOS Spectra}
\par This paper is based on the portion of the SPLASH survey that targeted the disk of M31: Keck II/DEIMOS line-of-sight velocity measurements of $\sim10,000$ stars \citep{Dorman2012, Dorman2013,Dorman2015}. The 600~lines~mm$^{-1}$ grating was used for Fall~2012 observations, and the 1200~lines~mm$^{-1}$ grating for the rest of the (earlier) observations. Not all stars observed in the SPLASH survey had HST photometry; some were selected from ground based imaging. The stars chosen for this study overlapped with those observed by HST in the PHAT survey and were targeted to encompass a range of stellar evolutionary points: young hot massive stars on the main sequence (MS), intermediate mass asymptotic giant branch (AGB) stars, and low mass red giant branch (RGB) stars. Milky Way Foreground (MWFG) stars dominate the region where $1<B-I<2$, so we avoid selecting stars from this region (see Section 3.1.1 in \cite{Dorman2015} and Figure \ref{fig:CMD} for more details). A typical velocity error for stars in this dataset is on the order of 20~km~s$^{-1}$. The target selection and data reduction process are described in \cite{Dorman2012, Dorman2013,Dorman2015}.

\subsection{HST PHAT Photometry}
\par The PHAT survey is comprised of 117~million stars in M31 that have been observed with HST/ACS and WFC3. We use optical photometry of $6,000$ stars that were bright enough for ground based spectroscopy and separated enough from nearby stars to be resolved and to avoid slit collision \citep{Williams2014}. Each star was observed in six filters, from the ultraviolet to the infrared: F275W, F336W, F475W, F814W, F110W, and F160W. The HST filter F275W corresponds to vacuum ultraviolet so does not have an equivalent ground based filter, but the remaining filters roughly correspond to the $U$, $B$, $I$, $J$, and $H$ filter, respectively. We make use of the F475W and F814W filters when separating stars into age bins (Section~\ref{sec:age}) and F110W and F160W when examining possible effects of reddening (Appendix~\ref{sec:RGB_red}).  

\subsection{DRAO HI Data}
\par \cite{Chemin2006} collected HI 21-cm line data of M31 using the Synthesis Telescope with the 26-m antenna at the Dominion Radio Astrophysical Observatory. The telescope has the capability of resolving features down to 58$\arcsec$~$\times$~58$\arcsec$/$\sin{(\delta)}$ in area. We use a subset of this catalog and only include data that correspond with the sightlines of the stellar data. As we explain in Section~\ref{sec:HI} in more detail, the HI spectrum can have multiple components with various possible origins (e.g. unresolved distinct HI disk structures within
the synthesized beam, extraplanar components, expanding clouds, and/pr the warping of HI disk). \cite{Chemin2006} choose to use the velocity that is the largest offset relative to M31's systematic velocity, and we make the same choice for this analysis. We experiment with the effect of this choice in Section~\ref{sec:HI}. 

\subsection{IRAM CO Data}
\cite{Nieten2006} observed the molecular CO transition of $\rm J=1 \rightarrow 0$ emission from November~1995 and August~2001 with the IRAM 30-m telescope. The observations were obtained in \it On-the-Fly \rm mode to obtain a total of 12 fields. The first set of fields was taken parallel to M31's minor axis. The field size was approximately 18$\arcmin$~$\times$~18$\arcmin$. The second fields were smaller in area, 9$\arcmin$~$\times$~9$\arcmin$, and were parallel to the major axis. The telescope scanned at a speed of $4\arcsec~s^{-1}$.  Because of the duration of the survey, the telescope was regularly moved to nearby targets to verify its pointing. See \cite{Nieten2006} Section 2 for more details about the focusing, observation plan, and data reduction. The CO emission is less likely to contain foreground contamination so is included in this analysis as a comparison to results relative to the HI. However, as seen in Figure \ref{fig:surveys}, the CO does not extend the full stellar or HI coverage so cannot be used as a comparison in the outer regions. 

\section{Asymmetric Drift as a Function of Stellar Age}\label{sec:AD_age}
\subsection{Age Groups} \label{sec:age} 
\par
\begin{figure}[h!]
\epsscale{1.2}
\plotone{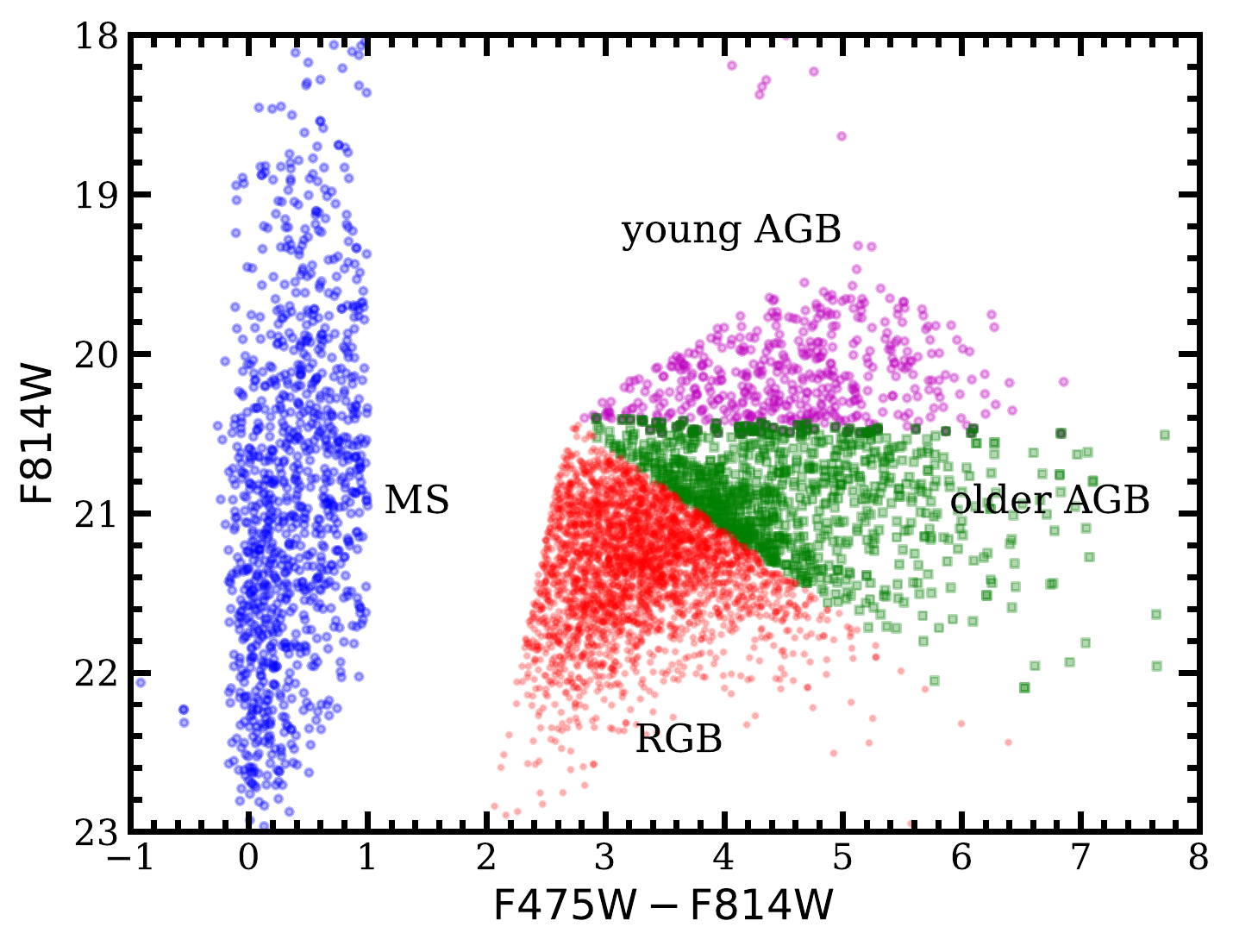}
\caption{CMD for the stellar sample used in this analysis using filters F475W and F814W from HST (roughly $B$ and $I$ equivalent). The age bin divisions are shown in the different colors. Blue points represent MS stars, purple points young AGB stars, green squares older AGB stars, and red points RGB stars. The average ages are 30~Myr for MS stars, 0.4~Gyr for the young AGB stars, 2~Gyr for the older AGB stars, and 4~Gyr for RGB stars. MWFG stars dominate the color space $1<\rm F475W-F814W<2$, so we avoid this region (see Section 3.1.1 in \cite{Dorman2015} for more details).\label{fig:CMD}}
\end{figure}

 The stellar sample is divided into four age bins, from shortest lived to longest lived: massive MS stars, intermediate mass young AGB stars, intermediate mass older AGB stars, and low mass RGB stars. The divisions are based on the star's location in color magnitude space, as seen in Figure~\ref{fig:CMD}. The CMD uses filters F475W and F814W from HST (roughly $B$ and $I$ equivalent). We adopt the average age of each subpopulation from \cite{Dorman2015}, who estimates the ages based on a simulated color-magnitude diagram with an age-metallicity relation derived from the RGB color distribution and assuming a constant star formation history. Since M31 has had multiple accretion events, it is unlikely to have had a constant star formation rate. However, a modeled CMD with a constant star formation history is valid in this analysis because we only want to place stars into broad age bins instead of identify their precise ages. This process yields average ages of 30~Myr for MS stars, 0.4~Gyr for the young AGB stars, 2~Gyr for the older AGB stars, and 4~Gyr for RGB stars.

When shown in $U-B$ color magnitude space, the MS star group shows three distinct populations that all have been categorized generally as MS: MS stars, blue loop stars, and yellow supergiants. Although these are three separate populations, all are young ($\ll\rm1~Gyr$). RCs for each group do not show noticeable differences between the rotation velocities of the three subpopulations, so the MS group is kept as one age bin to avoid smaller sample sizes.

\subsection{Line of Sight Kinematics}
\par
\begin{figure*}
\epsscale{1}
\plotone{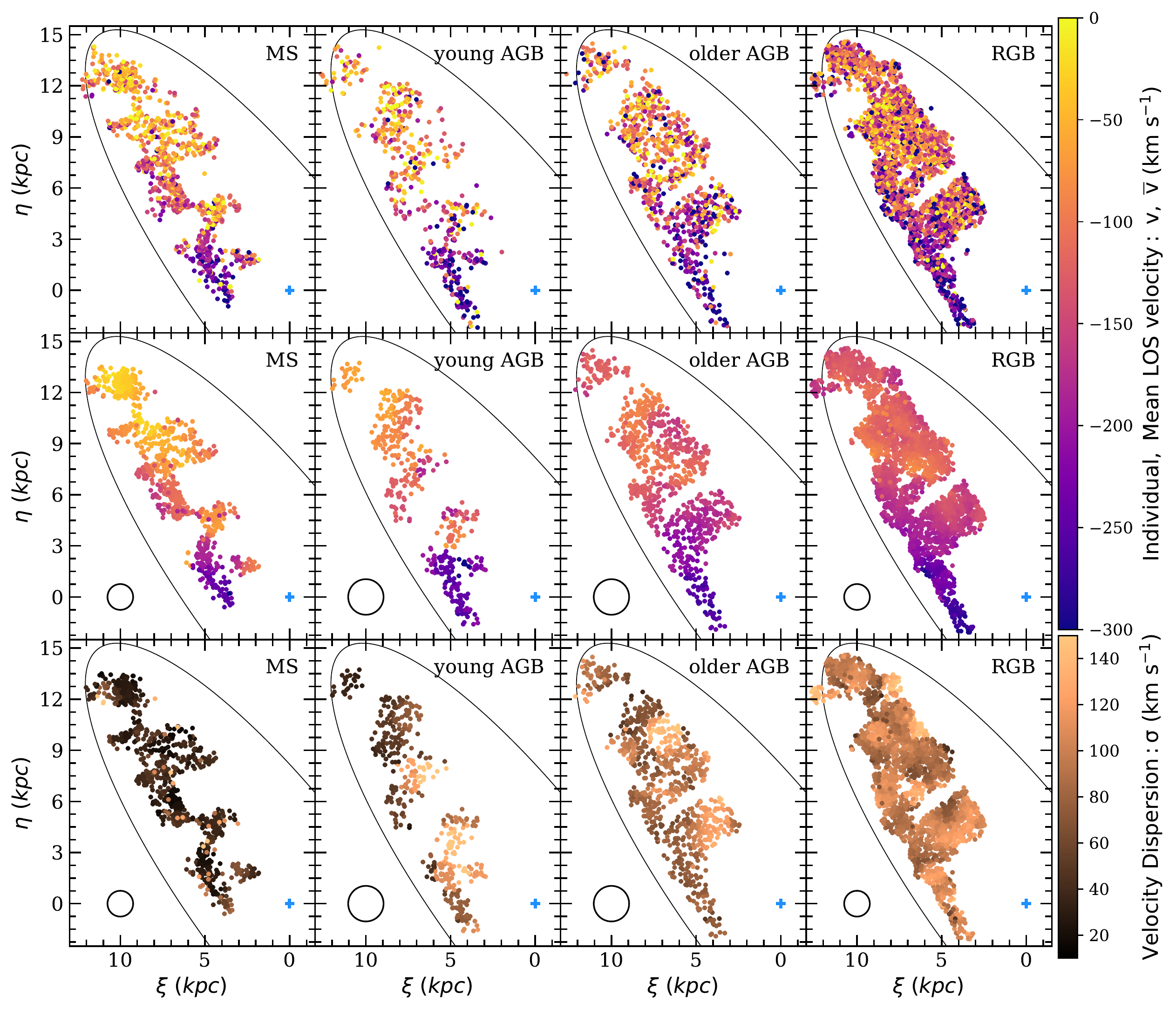}
\caption{Individual line-of-sight velocity (top row), locally averaged line-of-sight velocity (middle row), and local velocity dispersion (bottom row) as a function of location and stellar age. From left to right: massive MS stars, young AGB stars, older AGB stars, and RGB stars. Smoothing circles are used to calculate the weighted mean of the velocity and the weighted root mean square error of the velocity (velocity dispersion). For AGB stars, a size of 275\arcsec\ is used and a size of 200\arcsec\ for MS and RGB stars is used. The circles in each panel of the bottom two rows show the respective sizes of the smoothing circles. The blue cross marks the center of M31. The ellipse is for visual reference of the disk. Local velocity dispersion increases with stellar age while the overall range of LOS velocities decrease, as found in \cite{Dorman2015}. \label{fig:maps}}
\end{figure*}

 We examine the kinematics of each age group using the line-of-sight velocity obtained from the spectroscopy. The process of extracting velocities from the spectra is described in \cite{Dorman2015}. Figure~\ref{fig:maps} presents the position and individual line-of-sight velocity, average line-of-sight velocity, and velocity dispersion of each star for the four age bins. To average the velocities and calculate dispersion, we use a circle of 200\arcsec\ for MS and RGB and 275\arcsec\ for the less populated AGB bins. Each circle is centered on a star as in \cite{Dorman2015}. Only data from circles containing at least 15 stars are kept. In Figure~\ref{fig:maps} there is a clear increase in local velocity dispersion and a decrease in the overall range of line-of-sight velocity from the younger to the older stellar populations \citep{Dorman2015}. The increase in random motions with stellar age suggests that M31 has been dynamically heated in the past and that this heating perturbed stars in the disk. Continuous heating can explain the gradual monotonic increase across age with all four age bins. In the next subsection we further analyze the differences in the kinematics of each age bin.

\par The smoothing circle technique creates a dataset in which points are not independent of one another because the circles overlap. To test the effect of this, we perform velocity averaging in a different way to create a dataset of {\it independent\/} mean velocities as a function of sky position. Instead of using smoothing circles centered on each star, we place a rectangular grid on top of the individual velocity maps and derive a local average velocity and velocity dispersion from the stars within each grid cell. In this method, each resulting averaged velocity is independent of the other averaged values. To maximize the number of resulting data points while maintaining enough stars in each cell for proper statistics, a different grid size is used for each age bin: 150\arcsec, 216\arcsec, 198\arcsec, and 100\arcsec\ for MS, young AGB, older AGB, and RGB, respectively. A grid cell is required to contain a minimum number of stars in order to contribute to the map. This minimum number is four for MS, young AGB, and older AGB and ten for the more populated RGB group. The resulting velocity maps and RCs do not differ significantly from the ones derived with the smoothing circles. Naturally, the map based on the non-overlapping rectangular grid method is much sparser than the map based on the overlapping circles method.  Thus, we choose to use the averaging within overlapping circles analysis without the removal of outliers (with respect to the stars in each smoothing circle) for the rest of this paper so as to have the largest dataset possible.

\subsection{Rotation Curves}
\par Along with line-of-sight velocity, we compare the deprojected rotation velocities ($v_{\rm rot}$) of the stellar and gas populations. To deproject the averaged line-of-sight velocity into a rotation velocity, Equation~\ref{eq:v_rot} is used, where $v_{\rm sys}$ is the systematic velocity of M31 ($-300$~km~s$^{-1}$), $v_{\rm\star}$ is the averaged line-of-sight velocity of the star or gas at the same line of sight, and $PA_{\rm\star}$ is the position angle of the star or gas at the same line of sight based on its location in the disk.

\begin{equation} \label{eq:v_rot}
v_{\rm rot} = \pm \frac{v_{\star}-v_{\rm sys}}{\sin(i_{\rm TR})} \sqrt{1 + \frac{\tan^{2}(PA_{\star}-PA_{\rm TR})}{\cos^{2}(i_{\rm TR})}} 
\end{equation}

We use the tilted ring model from \cite{Chemin2006} to assign each star and gas line of sight a position angle ($\rm PA_{\rm TR}$) and inclination angle ($i_{\rm TR}$). This model is used to account for the warps in the HI disk. 
The tilted ring model consists of dividing the disk into different annuli, each with a corresponding PA and $i$. The rings have a width of 0.38~kpc, and the PA and $i$ for each ring used in this study can be seen in Table \ref{tab:table1}. \cite{Chemin2006} extend the model to $r=0.38$~kpc in the inner region and to $r=130$~kpc in the outer region, but that is beyond the extent of the data used in this analysis.
We use this model to place each star or gas sightline into a ring based on the deprojected distance from M31's center. We then assign a $\rm PA_{\rm TR}$ and a $i_{\rm TR}$ to the star or gas line of sight based on which ring it lies in. 

\begin{deluxetable}{ccc} 
\tablenum{1} \label{tab:table1}
\tablewidth{0pt}
\tablecaption{\cite{Chemin2006} Tilted Ring Model}
\tablehead{\colhead{Radius (kpc)} &  \colhead{adopted PA (\degr)} &  \colhead{adopted $i$ (\degr)}}
\startdata
4.95 & 33.6 & 63.7\\
5.33 & 33.9 & 65.9\\
5.71 & 35.4 & 68.1\\
6.09 & 36.4 & 69.7\\
6.47 & 36.6 & 72.0\\
6.85 & 36.5 & 73.5\\
7.23 & 36.4 & 74.3\\
7.61 & 36.8 & 74.6\\
7.99 & 37.3 & 74.5\\
8.37 & 37.7 & 74.3\\
8.75 & 38.0 & 74.3\\
9.13 & 38.3 & 74.4\\
9.51 & 38.7 & 74.8\\
9.90 & 39.0 & 75.2\\
10.28& 39.1 & 75.6\\
10.66& 39.0 & 76.1\\
11.04& 38.8 & 76.3\\
11.42& 38.6 & 76.4\\
11.80& 38.3 & 76.3\\
12.18& 37.7 & 76.0\\
12.56& 37.0 & 75.6\\
12.94& 36.4 & 75.1\\
13.32& 36.1 & 74.8\\
13.70& 35.9 & 74.5\\
14.08& 36.1 & 74.3\\
14.46& 36.6 & 74.1\\
14.84& 37.1 & 73.7\\
15.23& 37.3 & 73.7\\
15.61& 37.2 & 74.1\\
15.99& 37.1 & 74.7\\
16.37& 37.3 & 75.4\\
16.75& 37.7 & 75.5\\
17.13& 37.7 & 75.1\\
17.51& 37.5 & 74.3\\
17.89& 37.5 & 73.8\\
18.27& 37.5 & 73.4\\
18.65& 37.4 & 73.4\\
19.03& 37.5 & 73.6\\
19.41& 37.7 & 73.4\\
19.79& 38.3 & 73.2\\
20.18& 38.6 & 73.0\
\enddata
\end{deluxetable}

\par
\begin{figure*}
\epsscale{.9}
\plotone{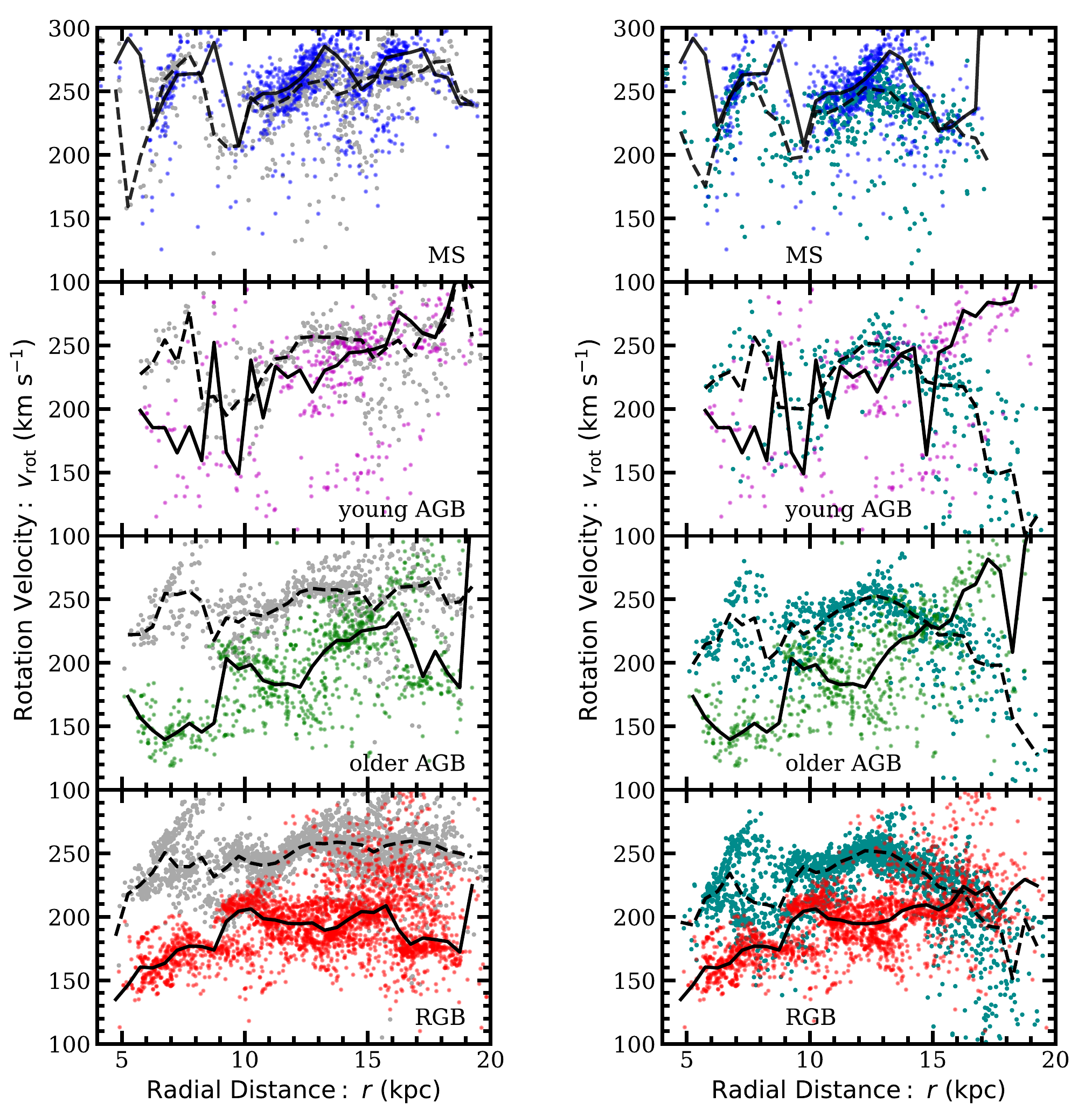}
\caption{Deprojected RCs for the four stellar age bins and the paired main component of the gas velocity. From top to bottom: massive MS stars, young AGB stars, older AGB stars, and RGB stars. In each panel on the left, the gray points represent the HI along the same line of sight as the stars, and the color points represent the stars. On the right, the dark teal points represent the CO along the same line of sight as the stars and the color points represent the stars. The lines represent the median rotation velocity in a 0.5~kpc bin. The solid line represents the stars and the dashed, the gas. A titled ring model was used to account for the warps in M31's disk when deprojecting the line-of-sight velocity into a rotation velocity, as used in \cite{Chemin2006}. The gap between the stellar RC and the gas RC, AD, increases monotonically with stellar age for both gas components. \label{fig:curves1}}
\end{figure*}

 The averaged line-of-sight velocity is used to create the RCs in Figure~\ref{fig:curves1}, which show the deprojected rotation velocity against the deprojected radial distance from the center of M31. 
Each star is paired with a HI and CO velocity measurement that is closest in right ascension and declination to the star's position.
The older populations show a decrease in the overall range of line-of-sight velocity compared to that of the younger populations (Figure~\ref{fig:maps}) and also show a clear transition to lower rotation velocities, as shown by the color points in Figure~\ref{fig:curves1}.  The RC for the HI along the stellar line of sight is represented by the grey points in the left panel of Figure~\ref{fig:curves1}, and the CO along the stellar line of sight RC is represented by the dark teal points in the right panel of Figure~\ref{fig:curves1}. The gas RCs are similar along all sightlines and do not vary regardless of whether the area of the disk is dominated by old or young stars. The extent of the CO data is smaller than the CO so does not extend to the outermost radii of the HI and stellar data. The young stellar age groups with higher rotation speeds more closely resemble the gas. Since the gas is collisional, it can be assumed to have a roughly circular orbital path. Interactions with the bar or with the spiral arms can perturb the gas, so this assumption is not necessarily correct in all circumstances but is adequate for measuring AD. The younger stars with lower dispersion match the rotation velocity of the gas more than the older stars do. When orbits are disturbed, the circular component of the orbital velocity becomes a smaller component of a star's overall motion. Thus, a lower rotation velocity can be an indicator of noncircular orbits. The gap between the gas RC and the stellar RC is a visual representation of AD: the stars lag behind the gas. For the RCs in Figure~\ref{fig:curves1}, the strongest component of the HI is used if the HI spectrum has multiple peaks. In Section~\ref{sec:HI} we discuss additional AD analysis with different choices of HI component. 

\par There is also scatter and substructure in the RC for each age bin for both gas components. When performing the velocity smoothing in Figure~\ref{fig:maps}, outliers within each smoothing circle are not removed. As an experiment, the analysis is repeated, this time we remove the outliers to examine how the RCs change: for each smoothing circle, stars with line-of-sight velocities that are different than the median of all the line-of-sight velocities within the circle by three times the width of the distribution are removed. Only centers that retain at least 15 stars in the smoothing circle are kept. Removing outliers eliminates some scatter in the RCs but preserves a majority of the substructure. We explore other possible sources of the scatter and substructure in Section~\ref{sec:RC_messy}.

\subsection{Asymmetric Drift}
The RCs in Figure~\ref{fig:curves1} show that there is an offset between the HI/CO and the stellar rotation velocity. We analyze this difference, the AD. AD is defined in Equation~\ref{eq:ad}. 

\begin{equation} \label{eq:ad}
v_{a} = v_{\rm rot,gas} - v_{\rm rot,\star}
\end{equation}

\begin{figure*}
\epsscale{1.2}
\plotone{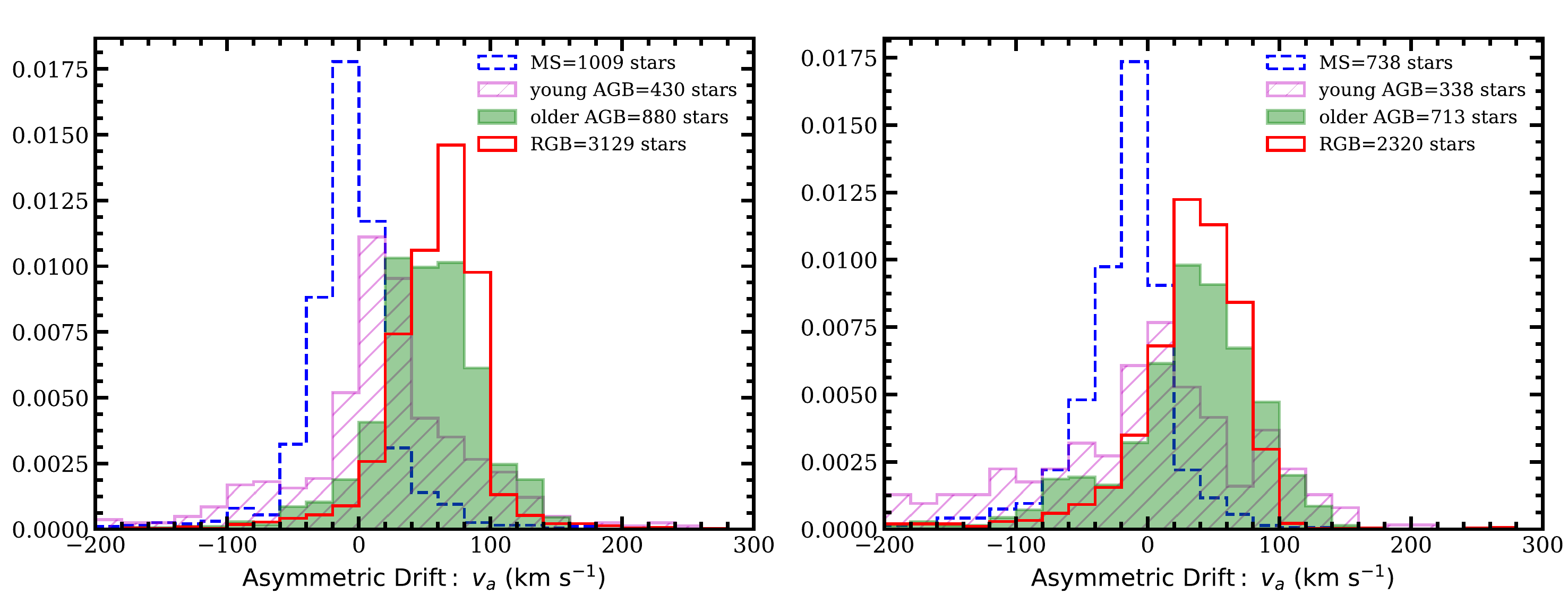}
\caption{Normalized distributions of AD (Equation~\ref{eq:ad}) for the four stellar age bins. In both panels, the histogram denoted by blue dashed line represent the massive MS stars, the ones with the purple stripes represents the young AGB stars, the green solid histograms represent the older AGB stars, and histograms denoted by the red line represent the RGB stars. The left panel shows the AD with respect to the HI and the right, AD with respect to the CO. The legend also contains the number of stars in each age bin. The magnitude of AD increases monotonically with stellar age. 
\label{fig:ad_hist}}
\end{figure*}

Histograms of AD for each age bin are presented in Figure~\ref{fig:ad_hist}. The left panel shows AD with respect to the HI, and the right shows AD with respect to the CO. The range of AD is largest for the young AGB stars. One result of this is interesting substructure, which is addressed later in the paper. In both panels, the lag becomes larger with stellar age: the peak of the distributions shifts away from zero as stellar age increases.
\par The AD measurements with respect to both gas components are offset from one another. The difference in AD measurement is likely a result of the fact that the velocities for each gas component are calculated differently. The HI line-of-sight velocities are derived from Gaussian fits on the most extreme velocity peak in the HI spectrum. The CO line-of-sight velocities are derived from the first moment of the spectrum. Gaussian fits result in higher rotation velocities than velocities derived from first moment maps \citep{Carignan1990}. 
\par
\begin{figure}[h!]
\epsscale{1.2}
\plotone{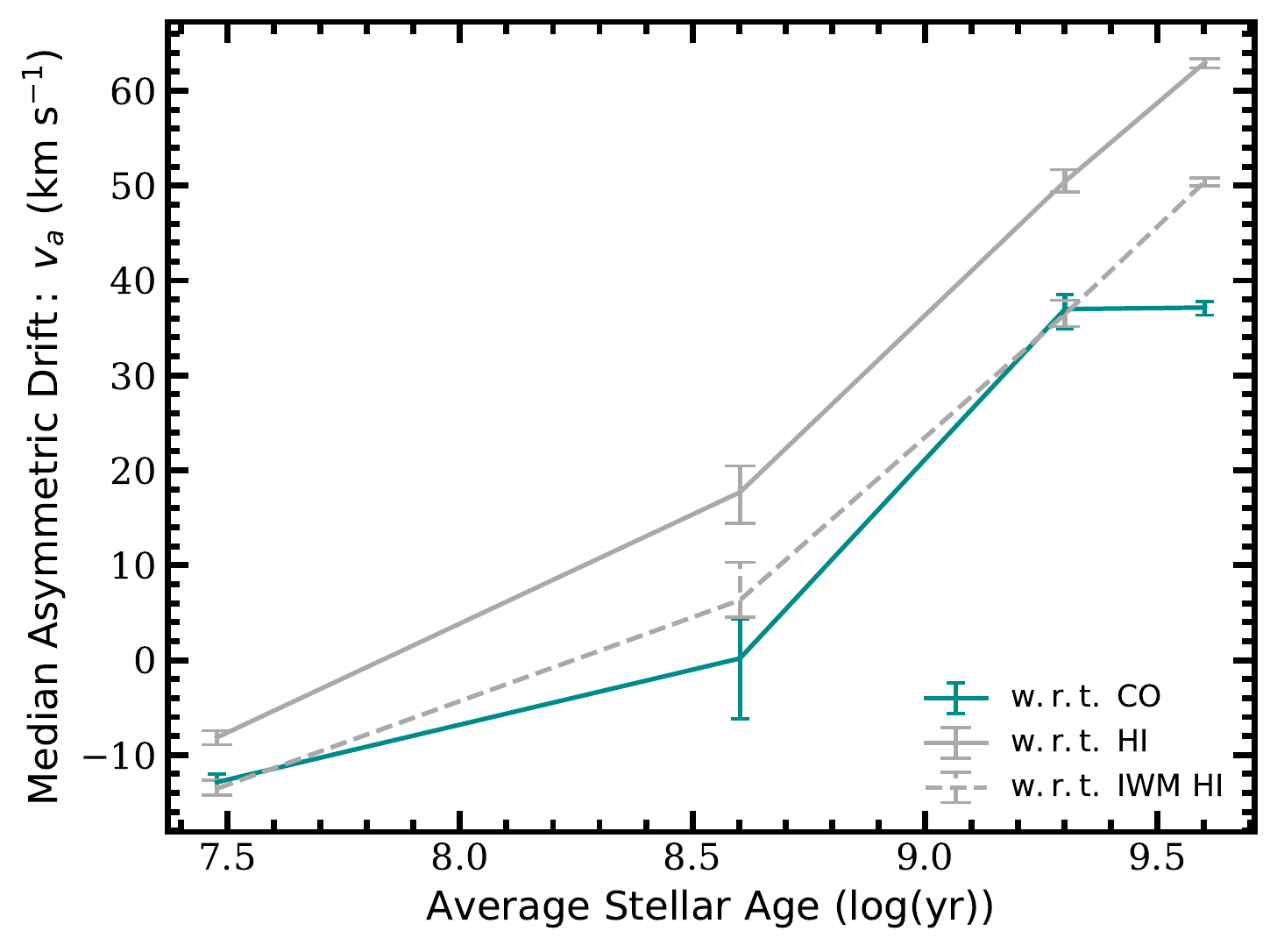}
\caption{Average stellar age against median AD. The upper and lower error bars represent the one sigma confidence level on the median. The teal represents AD with respect to the CO gas, and the solid gray line represents AD with respect to the HI gas, as shown in Figure \ref{fig:ad_hist}. The dashed gray line represents the AD with respect to the HI gas, where the velocity is intensity weighted (instead of determined by a Gaussian fit). The IWM method is more similar to how the CO velocities are determined. There is a monotonic increase in AD across every age bin, which suggests M31 has had relatively constant heating.}\label{fig:ad_transition}
\end{figure}

 For better comparison, we also derive the intensity weighted mean (IWM) of the HI line-of-sight velocities. To do this, we average the HI line-of-sight velocities across all HI peaks, which more closely resembles the first moment based CO velocities. The results of this analysis can be seen in Table \ref{tab:table2} and in Figure \ref{fig:ad_transition}. Table \ref{tab:table2} shows median AD values and the corresponding one sigma standard errors on the median. Figure \ref{fig:ad_transition} shows how the median AD values evolve with stellar age. The solid gray line shows the AD relative to the HI velocities that were derived by Gaussian fits, and the dashed gray line shows the AD relative to the HI IWM velocities. The difference between the HI IWM AD measurements and CO AD measurements is smaller than the AD measurement relative to the HI Gaussian fit velocities. The remaining small differences could be caused by different resolutions, gas properties, and the fact that while the two velocity measurement methods, first moment versus IWM are more similar than before, they are not identical. Furthermore, we see that the CO gas rotation speeds better match the rotation speeds of the young MS stars, which is what is expected based on the fact that young stars are born out of molecular gas. In the rest of the paper, we use HI line-of-sight velocities based on Gaussian fits to the most extreme velocity peak in the spectrum because the complexity of the HI spectrum makes the IWM an inferior measure of the rotation speed. 
\par Surprisingly, the MS population shows a negative median AD value, suggesting the HI and the CO gas lag behind the stars. This implies the young stars are more settled into the circular velocity of the galactic potential than the collisional gas. Young stars are born in gas that is denser than the gas in this analysis so it is most likely that using HI or CO ($\rm J=1 \rightarrow 0$) as a proxy for circular motion is not perfect. Figure \ref{fig:MS_rcs} shows that not all of the younger stars are leading the gas.  In particular, there are groups of stars at 13-15~kpc that trail the gas like the other age groups. The explanation for this is not apparent.
The two oldest age bins lag behind the stars on average by 20\% and up to 40\%, which is greater than the AD that is seen in the MW and other local galaxies \citep{Ciardullo2004,Olling}. This suggests M31 has had a particularly violent heating history. 

\begin{deluxetable*}{cccc} 
\tablenum{2} \label{tab:table2}
\tablewidth{0pt}
\tablecaption{Median AD Values}
\tablehead{\colhead{Stellar Group} & \colhead{AD w.r.t HI $(\rm km\ s^{-1})$} &      \colhead{AD w.r.t HI IWM $(\rm km\ s^{-1})$} &  \colhead{AD w.r.t CO $(\rm km\ s^{-1})$}}
\startdata
MS & $-8.15^{+0.74}_{-0.72}$ & $-13.54^{0.68}_{0.88}$ & $-12.86^{+1.38}_{-0.86}$\\
young AGB & $17.69^{+3.29}_{-2.78}$ & $6.32^{1.79}_{3.99}$ & $0.18^{+6.38}_{-4.17}$\\
older AGB & $50.43^{+1.09}_{-1.26}$ & $36.45^{1.30}_{1.45}$ & $36.99^{+2.09}_{-1.55}$\\
RGB & $62.97^{+0.59}_{-0.40}$ & $50.43^{0.46}_{0.39}$ & $37.15^{+0.79}_{-0.64}$\
\enddata
\end{deluxetable*}

\par Examining the transition to decreased rotational velocity with stellar age can give us detailed insight into the heating history of M31. Previous studies on the formation of M31 have not considered AD of different stellar populations. In theory, a smooth transition between the age bins suggests the heating was continuous. On the other hand, a sudden jump in lag between the age bins instead suggests that there was one event or sporadic events that heated the disk.  
In this analysis, we see an increase in lag at all age bins, as shown in Figure \ref{fig:ad_transition}. 
This monotonic increase of AD suggests M31 has had relatively continuous accretion: there have been mergers before and after larger events, like the creation of the GSS \citep{Ferguson2016,Sadoun2014} or the event that caused the star formation burst that occurred 2~Gyr ago \citep{Williams2017,Williams2018,DSouza2018}. While this analysis supports that multiple and relatively frequent mergers are needed to explain the trend of increasing AD with stellar age in M31, we cannot determine whether minor or major mergers are more important for creating this phenomenon. A comparison to simulations will be the subject of future work and will better answer the question of whether a major merger or minor mergers can recover the current dynamics of M31's disk.  

\par Additionally, the shape of the velocity ellipsoid is beyond the scope of this paper but will be the subject of future analysis. This future analysis will include an extended sample of stars with greater azimuthal coverage but coarser age resolution. Measuring the shape of the velocity ellipsoid will constrain tangential anisotropy and will help determine the accuracy of the Jean's equation to model AD. Preliminary work with the asymmetric Jean's equation was done in \cite{DormanThesis}, and the predicted AD values are consistent with our measured values.

\section{Exploring the Substructure in the RCs (and Resulting Scatter in the Asymmetric Drift)}\label{sec:RC_messy}

\par
\begin{figure*}
\epsscale{1.1}
\plotone{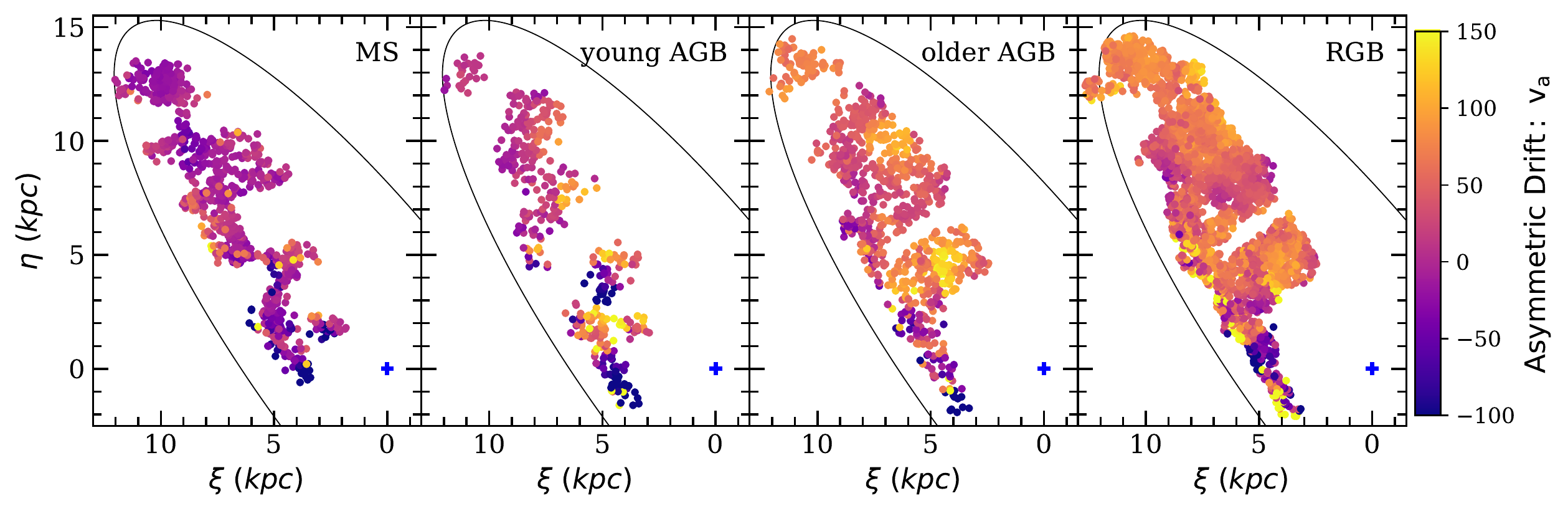}
\caption{Maps of AD for the four age bins. Left to right: massive MS stars, young AGB stars, older AGB stars, and RGB stars. Color represents AD. The center of M31 is marked with the blue plus symbol in each panel. The ellipse is for visual reference. AD increases with stellar age. \label{fig:ad_maps}}
\end{figure*}

 As presented in the previous section, there is a clear correlation between lag and stellar age. In addition to this trend, the RCs in Figure~\ref{fig:curves1} show substructure and scatter that cannot immediately be explained. In this section, we explore different possible sources that could be causing the messiness, and we examine the effects on AD measurements. Figure \ref{fig:ad_maps} shows maps of the four stellar age groups, and color represents AD values. This plot will be referred to throughout this section. Because the CO data does not extend the full stellar disk and does not contain peak multiplicity information, we limit the following analysis to the HI gas.

\subsection{Halo Star Contamination}
Because the disk is thick but not infinite, there is a risk that a sample of this kind could contain halo stars that are at the same sightlines as our data but beyond the finite disk scale height. Halo stars are old, and they exhibit lower rotation velocities than older stars in a disk. Including halo stars in our sample would skew our measured AD value. \cite{Dorman2013} decompose the bulge, disk, and halo of M31 using the luminosity function of bright stars from the PHAT survey, radial velocities of RGB stars from the SPLASH survey, and surface brightness profiles. Using these three criteria, they find that our sample is dominated by disk stars with negligible the halo and bulge star contamination. Furthermore, while having halo stars in our sample would increase the RGB AD, it would not affect the AD measurements for the other stellar age bins. We see that AD increases monotonically with age across all four age bins, so it is unlikely the high AD for the RGB group is caused by the presence of halo stars in that particular subsample.
 
\subsection{Geometrical Effects} \label{sec:geometry}
\par
\begin{figure*}
\epsscale{.9}
\plotone{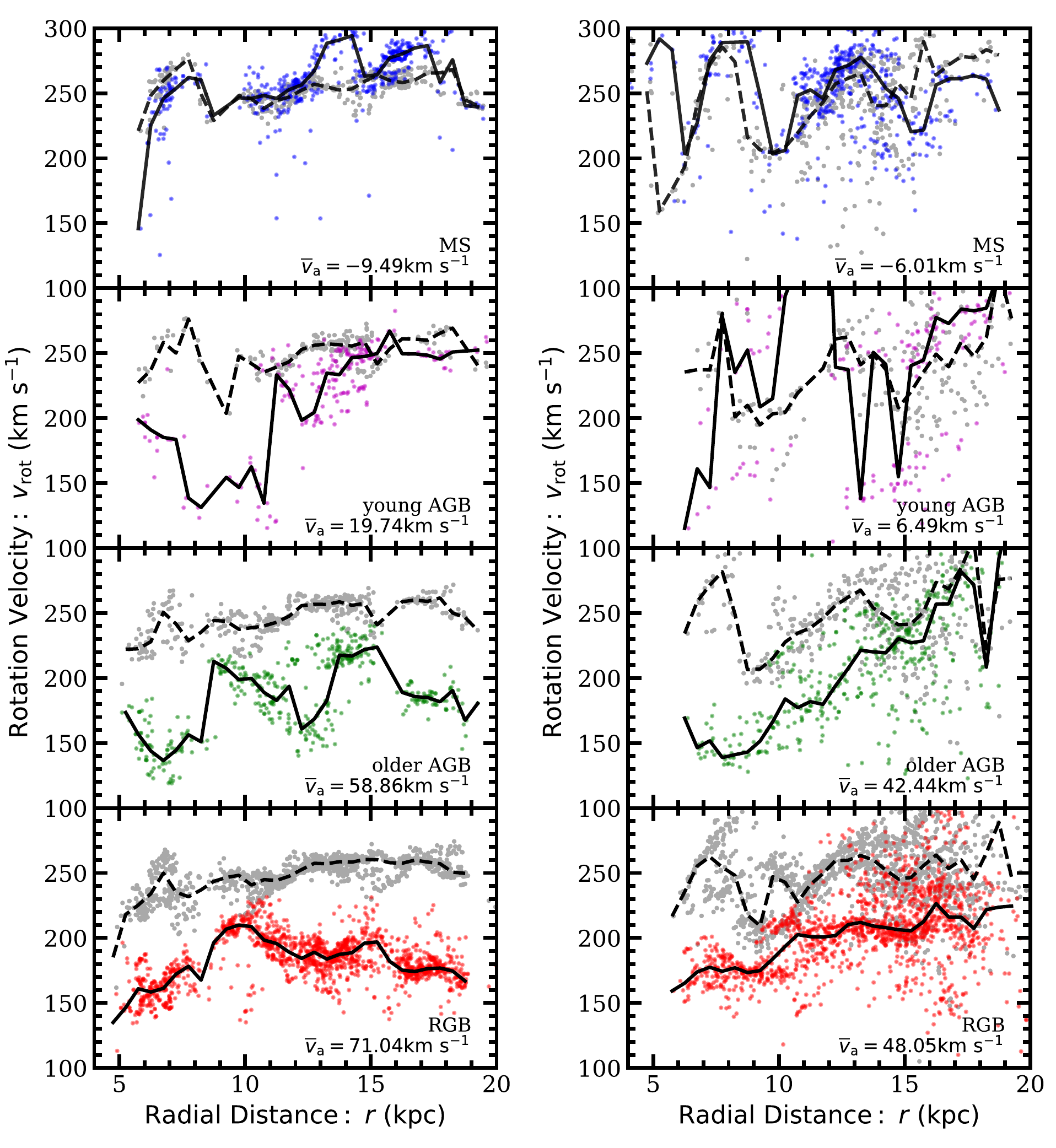}
\caption{Left panel: the RCs for sightlines where the deprojection factor from Equation~\ref{eq:v_rot} is lower than the median factor for each age group. These sightlines represent stars in the wedge denoted by the aqua lines in Figure~\ref{fig:surveys}. The right panel is the RCs for sightlines where the deprojection factor is above the median for the age group. From top to bottom: massive MS stars, young AGB stars, older AGB stars, and RGB stars. In each panel, the gray points represent the HI and the color points represent the stars. The lines represent the median rotation velocity for 0.5~kpc bins. The solid line is for the stars and the dashed for the gas. Each panel shows the median AD values. The left panels show less scatter and a clearer trend of increasing AD with stellar age.\label{fig:factor_rcs}}
\end{figure*}

 There is a possible geometrical effect resulting from the tilted ring model (Equation~\ref{eq:v_rot}) that could cause some of the the RC scatter in Figure~\ref{fig:curves1}. The deprojection factor is the term inside the square root in Equation \ref{eq:v_rot}, and it depends on position angle. This term varies from one to infinity depending on where the star is in the disk. If the star lies on the major axis, this factor is at a minimum. Stars with small deprojection factors lie in the wedge area denoted by the aqua lines in Figure~\ref{fig:surveys}. If however, the star is on the minor axis, the deprojection term approaches infinity, regardless of the inclination: on the minor axis $\rm PA_{\star}-PA_{\rm TR}=90$, so the tangent will be undefined. This drastic increase in the projection factor amplifies the inability of the tilted ring model to perfectly represent the disk. We separate sightlines for each age bin into two categories: the first has deprojection factors less than the median of the age bin, and the second has factors greater than the median. The left panel of Figure~\ref{fig:factor_rcs} shows the RCs that correspond to sightlines with lower deprojection factors, and the right panel shows the RCs that correspond to sightlines with higher deprojection factors. The RCs that correspond to sightlines with lower deprojection factors are flatter and contain less scatter than those corresponding to sightlines with a greater deprojection factor. This is especially clear in the HI RCs, which do not depend on stellar age. Because Equation~\ref{eq:v_rot} has no age dependence, each age bin is affected by the behavior of the deprojection factor equally. The trend of monotonically increasing AD with stellar age is clearer in the RCs that correspond to low deprojection factors. It is even stronger in these panels, as compared to the right panels, in which this trend is overshadowed by the scatter in the RCs. Furthermore, both the gas and the stellar line-of-sight velocity are deprojected by the titled ring model, so the model should not influence our measurements of AD. Figure \ref{fig:ad_maps} shows there is no strong azimuthal preference for higher or lower AD. There is a clustering of negative AD values with a large magnitude in the right most panel; these high AD values could be coincident with the bar. Most of the scatter in the RCs can be contributed to the tilted ring model being imperfect, but the source of the greater substructure in the RCs is still unclear. The RCs for the sightlines with low deprojection factors reveal the dramatic substructure in the older AGB population. Perhaps there are old AGB stars along the same line of sight as gas left over from recent star formation that is causing the shape of the RC. This substructure could also be a result of minor mergers that disrupted this population of stars.

\par The right panel of Figure \ref{fig:factor_rcs} shows the tilted ring model is not sufficient to describe the complexity of M31's disk structure. We construct RCs using a infinitely thin disk model, as described by Equation \ref{eq:v_rot_infinite} to see if perhaps a simpler model decreases the scatter in the RC. Unlike the titled ring model, each star is given the same $\rm PA_{TR}$ and $\rm i_{TR}$. These values are the average of the $\rm PA_{TR}$ and $\rm i_{TR}$ ($\rm PA_{TR, avg}$ and $\rm i_{TR, avg}$) values used in Figure \ref{fig:curves1}. 

\begin{equation} \label{eq:v_rot_infinite}
v_{\rm rot} = \pm \frac{v_{\star}-v_{\rm sys}}{\sin(i_{\rm TR, avg})} \sqrt{1 + \frac{\tan^{2}(PA_{\star}-PA_{\rm TR, avg})}{\cos^{2}(i_{\rm TR, avg})}} 
\end{equation}

\begin{figure}[h!]
\epsscale{1}
\plotone{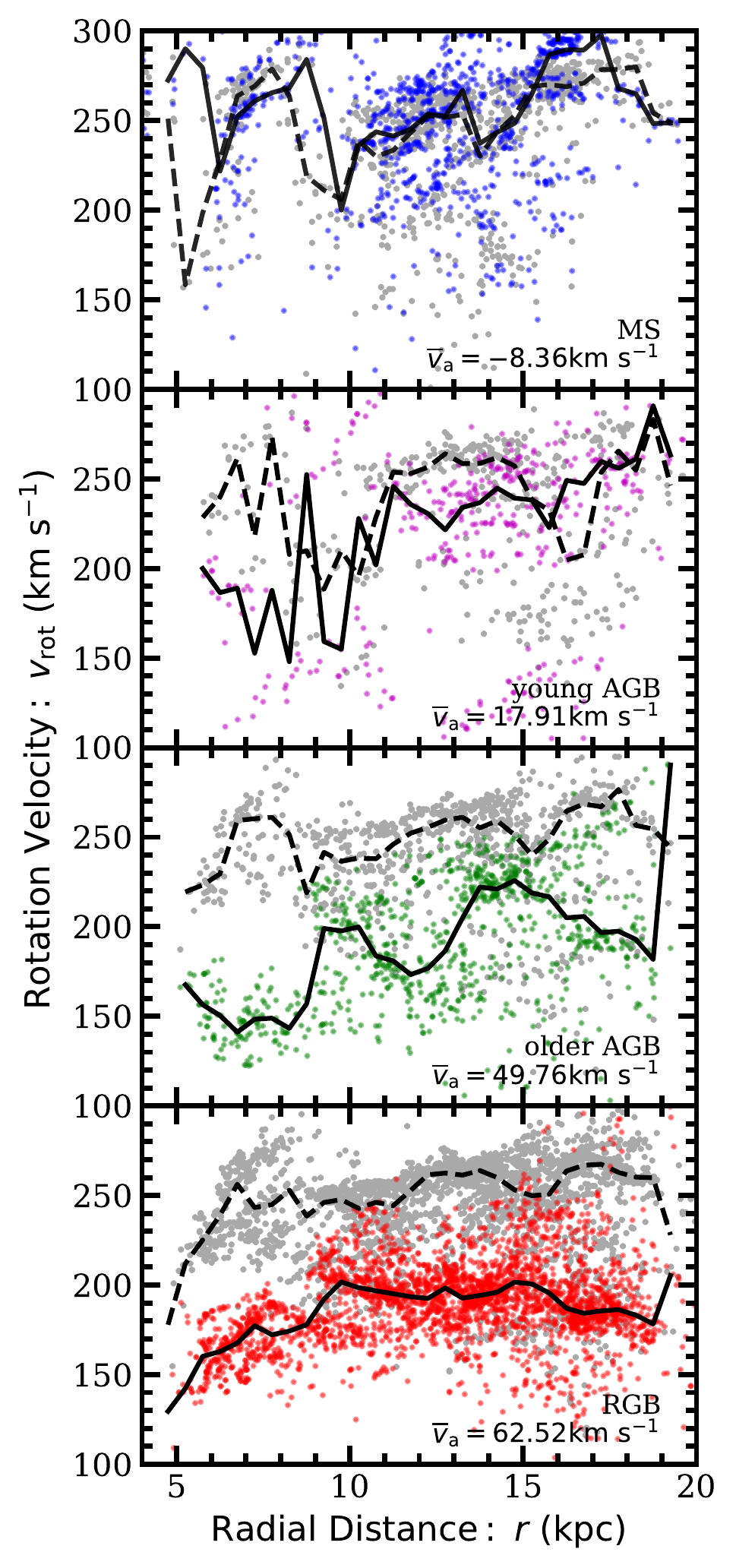}
\caption{Deprojected RCs for the four stellar age bins and for the main component of the HI along the corresponding line of sight using an infinitely thin disk model instead of the tilted ring model. From top to bottom: massive MS stars, young AGB stars, older AGB stars, and RGB stars. In each panel, the gray points represent the HI and the color points represent the stars. The lines represent the median rotation velocity for 0.5~kpc bins. The solid line is for the stars and the dashed for the gas. Each panel shows the median AD values. Like in Figure \ref{fig:curves1}, there is scatter and substructure in the curves, which shows a simpler model does not better represent the geometry of M31's disk. \label{fig:curve_infinite}}
\end{figure}

The RCs using this model are shown in Figure \ref{fig:curve_infinite}. The resulting curves perhaps have less substructure than those in Figure \ref{fig:curves1} but not by a significant amount. This suggests that the geometry of M31's disk cannot be explained by a simple model either and also suggests that the substructure. Despite the inadequacy of the tilted ring model and infinitely thin disk model, we see that AD increases monotonically as a function of stellar age. Since this trend exists in both models, we believe that this is a real phenomenon.   

\subsection{The Multiplicity in the HI Spectrum} \label{sec:HI}

\par \cite{Chemin2006} observe the HI gas in M31 and find inner and outer HI warps. A single line of sight can thus pass through several HI clouds, causing there to be more than one peak in the HI spectrum. \cite{Chemin2006} do not limit their data analysis to a single or double peak fit so as to be able to account for more than two peaks in the spectrum. They find the multiplicity of HI ($N_{\rm HI}$) can range from one to five. 
\par
\begin{figure*}
\epsscale{1.1}
\plotone{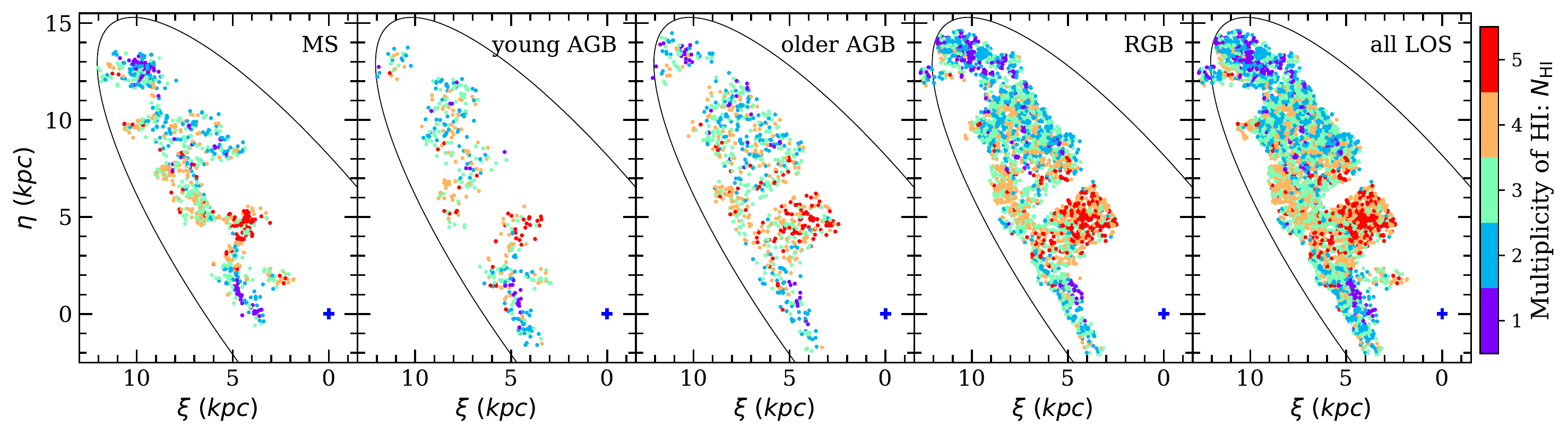}
\caption{Multiplicity of HI maps for the four age bins. Left to right: massive MS stars, young AGB stars, older AGB stars, and RGB stars. Color represents the number of peaks in the HI spectrum along the line of sight. The center of M31 is marked with the blue plus symbol in each panel. The ellipse is for visual reference. The sightlines with multiple components in the HI spectrum are more likely to be in inner regions and may be coincident with the bar. \label{fig:n_maps}}
\end{figure*}

 One result of this complexity is that the positions of stars can be measured inaccurately even when using a tilted ring model: a star and HI line of sight can be incorrectly assigned to the same radial bin, making substructure in the RCs.  Figure~\ref{fig:n_maps} shows maps of the stars in this sample, and the color represents the number of components in the HI spectrum at that line of sight. Inner radii are more likely to have multiple components than outer radii. This is consistent with findings from \cite{Chemin2006}: the HI disk is highly inclined at large radii, and the outer warp is projected along the line of sight onto inner regions, giving the illusion that there are multiple warps at small radii. The complexity in the inner regions is coincident with the location of the bar, which causes perturbations in the gas structure and dynamics via disruption and/or collision of gaseous clouds in warped crowded orbits and high speed clouds along orbits aligned with the bar. This also disrupts the HI inner ring, which causes structures at $r\sim~\rm2.5~kpc$ and $r\sim~\rm5~kpc$ \citep{Chemin2006}. These two sources of complexity in the HI line-of-sight velocities, (1) the outer warp leading to the projection of outer gas onto gas at smaller radii and
(2) interactions with the bar, could cause substructure in the RCs. Comparing Figure \ref{fig:n_maps} and Figure \ref{fig:ad_maps}, we see the most complex sightlines, as denoted by five peaks in the HI spectrum, appear to also be the location of the greatest AD. The interactions with the bar that are causing the substructure in the RC could also be a source of the high magnitude of AD in this region. Thus, the gas and the stars might be being perturbed by resonances associated with the bar.

\par The multiple peaks in the HI spectrum means a line of sight has multiple HI velocities. Typically, the velocity corresponding to the strongest peak (the velocity is that is largest in magnitude relative to the velocity of the system) is used in analyses, as we have done in this paper. We experiment with how this choice affects the RCs by comparing two velocities: that of the strongest peak and that of peak that corresponds to the velocity that is closest to the star's velocity at the same line of sight. The RCs do not differ in a noticeable way, so there is no trend between this choice of HI velocity and the stellar velocity dispersion (or AD). The RCs for this analysis can be seen in Appendix \ref{sec:HI_appendix}.
\par
\begin{figure*}
\epsscale{.9}
\plotone{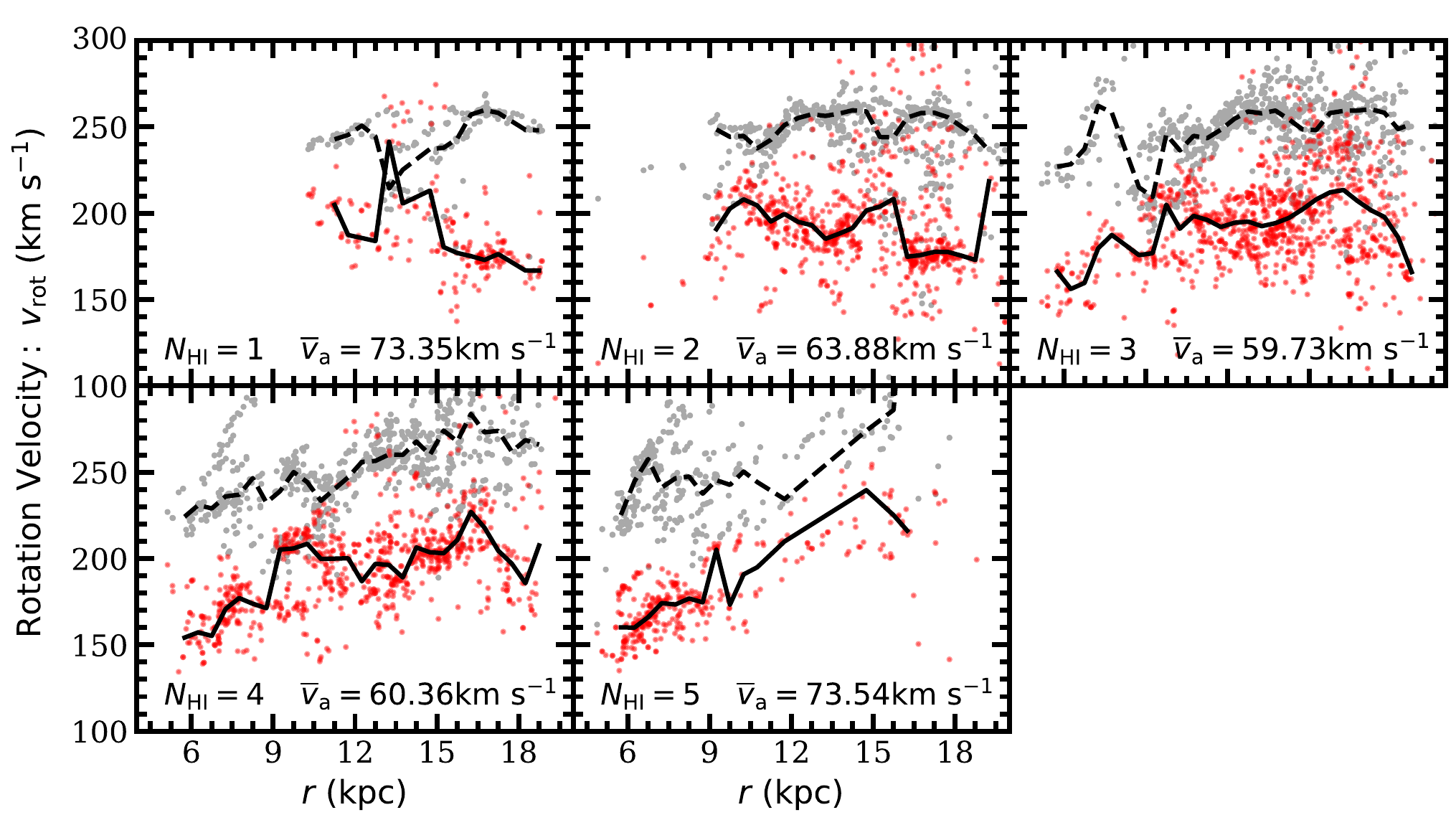}
\caption{RCs for the RGB stars grouped into subsamples based on the number of peaks in the HI spectrum along the corresponding sightlines. The number of peaks range from one to five. The red points represent the star velocities and the gray, the HI. The lines represent the median rotation velocity for a 0.5~kpc bin. The solid line is for the stars and the dashed the gas. Each panel gives the median AD. There is a clear lack of stars that correspond to a line of sight with one peak in the HI spectrum in the inner disk region. This region is well populated by stars along sightlines with five peaks in the HI spectrum, indicating that the inner $\rm10~kpc$ of the HI disk is messier than the outer radii. The stars still lag behind the gas for each subsample regardless of the number of components in the HI spectrum at that line of sight. \label{fig:RG_rcs}}
\end{figure*}

 We also examine if the number of components in the HI spectrum is correlated with substructure in the RCs. Figure~\ref{fig:RG_rcs} shows the RCs for the RGB stars grouped into subsamples based on the number of peaks in the HI spectrum along the corresponding sightlines, from one peak to five peaks. While there is a clear lack of data points in the first $\rm10~kpc$ whose line of sight correlates to an HI spectrum with a single peak, there is substructure and scatter regardless of the number of HI components. This is consistent with Figure~\ref{fig:n_maps} and findings by \cite{Chemin2006} that the inner radii of the HI disk have a more complicated shape. In the warped inner region, it is especially difficult to account for the geometry of the disk even with a tilted ring model \citep{Sadoun2014}. The median AD value varies throughout the subgroups but is surprisingly almost the same for the group with $N_{\rm HI}=1$ as for the group with $N_{\rm HI}=5$. Furthermore, the stars still lag behind the gas regardless of the number of peaks in the corresponding HI spectrum. We see similar findings in all age bins; the RCs for the shorter lived MS stars, young AGB stars, and older AGB stars can be seen in Appendix \ref{sec:HI_appendix} .

\par
\begin{figure*}
\epsscale{.9}
\plotone{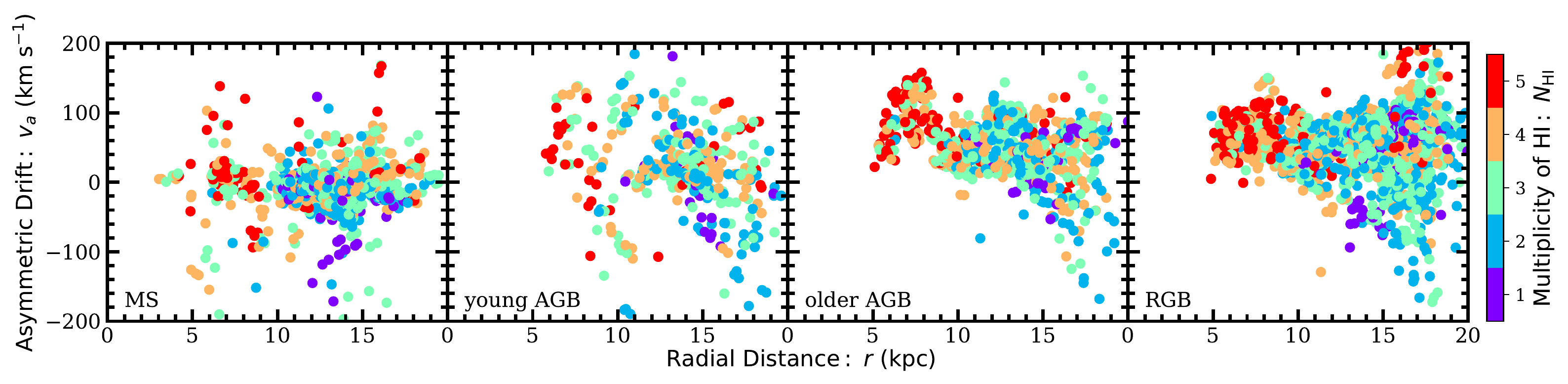}
\caption{AD as a function of radius. The color is the number of peaks in the HI spectrum. The purple points represent sightlines that have one peak in the HI spectrum, blue represents those that have two peaks, green represents three peaks, orange represents four peaks, and red represents five peaks.  From left to right: the massive MS stars, the young AGB stars, the older AGB stars, and the RGB stars. The range of AD increases at outer radii for the older three populations. \label{fig:r_lag_n}}
\end{figure*}

 Figure~\ref{fig:r_lag_n} shows the radial distance, AD, and number of components in the HI spectrum along the line of sight for each star in the four age bins. For the AGB and RGB stars, there is a greater width of the distribution of lag at large radii ($r~\rm>~15~kpc$). This is counter to what is expected from the trend seen in Figure~\ref{fig:RG_rcs}: since the HI spectrum is better behaved at outer radii, we would expect there to be less of a spread in lag at large radii and instead see a wider distribution of AD at smaller radii where the HI has multiple components in its spectrum. This discrepancy could be signs of an additional outer warp that does not present itself in the number of components in the HI spectrum.
\par
\begin{figure*}
\epsscale{.9}
\plotone{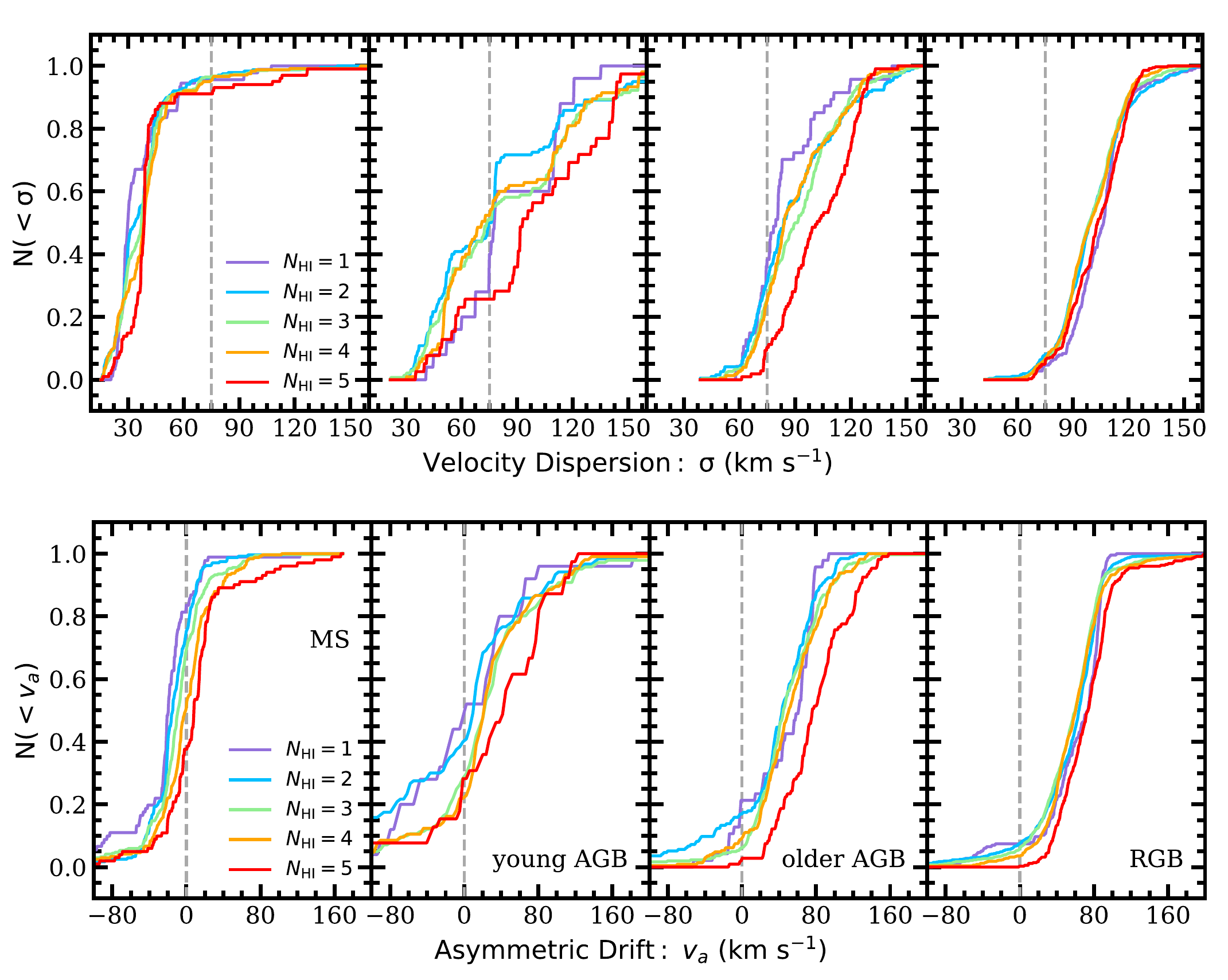}
\caption{Cumulative histograms for velocity dispersion (upper panels) and AD (lower panels) for subsamples based on the number of peaks in the HI spectrum along the corresponding sightlines. From left to right: massive MS stars, young AGB stars, older AGB stars, and RGB stars. The purple line represents sightlines that have one peak in the HI spectrum, blue represents those that have two peaks, green represents three peaks, orange four peaks, and red represents five peaks. The vertical dotted line marks $\rm \sigma=75~km~s^{-1}$ and $\rm v_{a}~=~0~km~s^{-1}$ for reference. Stars along sightlines with five components in the HI spectrum have greater velocity dispersion and AD than those with fewer components. Despite the number of peaks in the HI spectrum, longer lived stars have greater velocity dispersion and AD.} \label{fig:disp_lag_n_hists}
\end{figure*}

 We also look for possible correlations between the complexity of the line of sight as indicated by the HI and the velocity dispersion or AD. The top panel of Figure~\ref{fig:disp_lag_n_hists} shows cumulative histograms of the velocity dispersion for each age group. The bottom panel shows the cumulative AD histograms for the subgroups. The age groups are further divided into subpopulations based on the number of peaks in the HI spectrum along the line of sight. 
In general, the sightlines corresponding to multiple peaks in the HI spectrum have a greater magnitude of AD and velocity dispersion than the sightlines with a single HI component. Similarly, sightlines with five components in the HI spectrum always have greater velocity dispersion and AD magnitude than the other sightlines that correspond to fewer HI peaks. This is especially clear for the MS age group: the fraction of stars with $v_{a}~\leq~0$ decreases as the number of HI components increases. 
The difference between the multiplicity subgroups in the RGB age bin is smallest; there is less of a correlation between multiplicity in the HI spectrum and lag (or velocity dispersion) for longest lived populations. The dependence on HI multiplicity is most significant in the young AGB stars. Figure~\ref{fig:disp_lag_n_hists} also has vertical reference lines to highlight AD and velocity dispersion increase with stellar age despite the number of peaks in the HI spectrum.


\section{Conclusion}\label{sec:Summary}
In this study:
\begin{itemize}
  \item We use observations from the PHAT and SPLASH surveys of M31 along with 21-cm line measurements to compare the stellar rotation velocities of $\sim 6,000$ stars, which we divide into four aged populations based on color (30~Myr, 0.4~Gyr, 2~Gyr, and 4~Gyr), to the rotation velocities of two gas components (CO and HI).
  \item We show that AD is a function of stellar age: the youngest stars lead the HI and the longer lived populations lag behind the gas. The lag between the stars and the gas increases monotonically across each age bin, which suggests M31 has experienced continuous heating events. The median AD values for each stellar population with respect to the HI are $-8.15,\ 17.69,\ 50.43,\ \rm and\ 62.97$.
  \item We present substructure in the RCs and explore possible sources: geometrical effects and multiplicity in the HI spectrum. A significant portion of the scatter in the RCs is caused by a geometrical effect of the tilted ring model that is used to deproject line-of-sight velocity into a rotation velocity. This demonstrates that the tilted ring model is not sufficient to describe the multiple warps in the disk of M31. Despite the limitations this puts on our ability to perfectly model the disk and its dynamics, we show that the underlying trend of monotonically increasing AD with stellar age is a true phenomenon.
\end{itemize}
\par ACNQ and PG are grateful for support from the National Science Foundation grant AST‐1412648 and NASA/STScI grants GO‐12055, GO-14268, and GO-14769. The research of LC is supported by the Comit\'e Mixto ESO-Chile and the DGI of Universidad de Antofagasta. We would also like to thank the anonymous referee for the helpful comments that added to the depth of this paper. We appreciate the significance of Mauna Kea to the indigenous Hawaiian community and acknowledge we are fortunate to be able to use observations taken from its summit.


\newpage 
\clearpage
\bibliographystyle{apj}
\bibliography{Mendeley} 


\appendix 
\subsection{The Multiplicity of the HI spectrum} \label{sec:HI_appendix}
\begin{figure}[h!]
\plotone{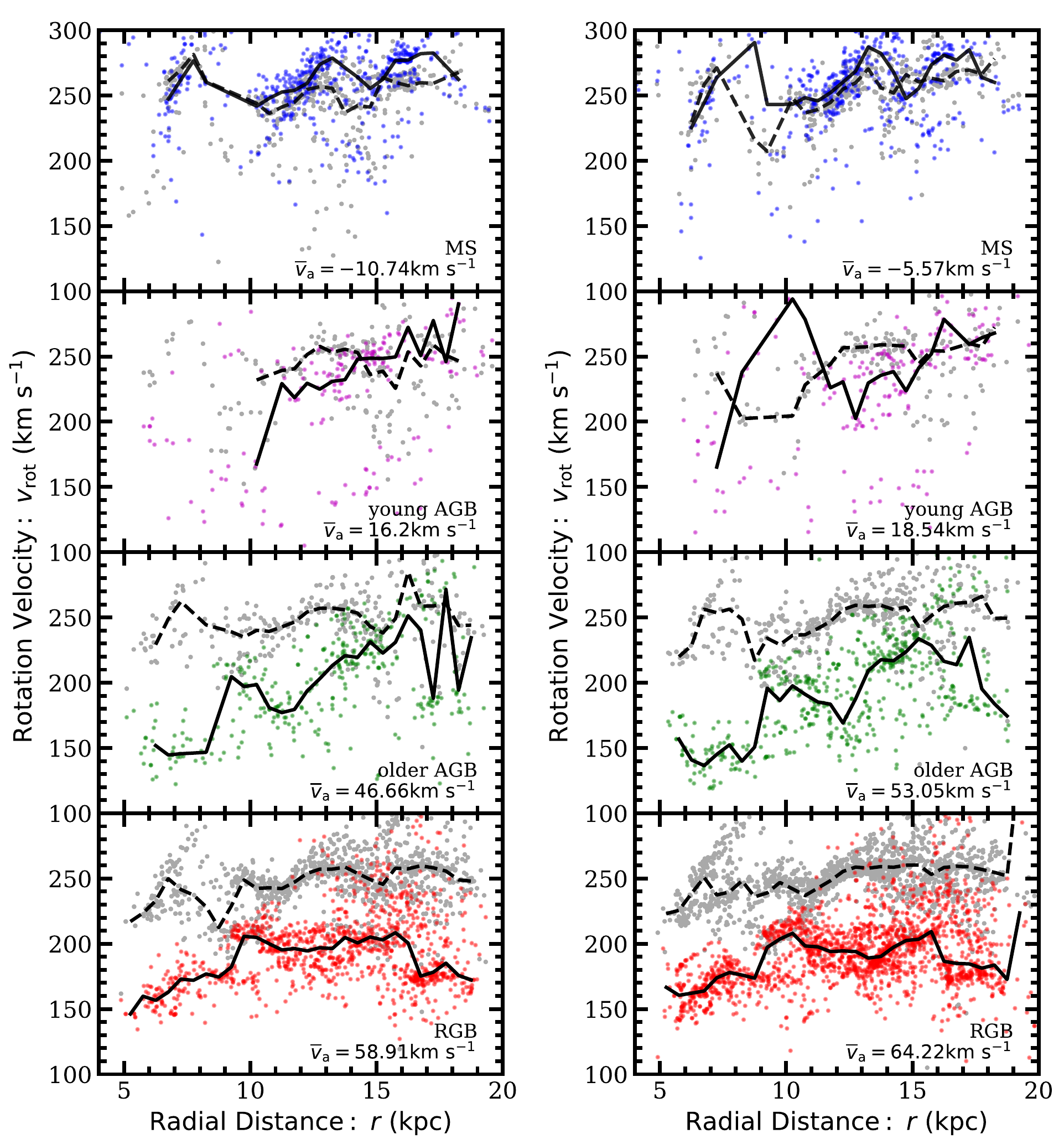}
\caption{Left panel: the RCs for sightlines in which the strongest HI component is also the peak associated with the velocity closest to that of the star. The right panel is the RCs for sightlines where this is not the case. From top to bottom: massive MS stars, young AGB stars, older AGB stars, and RGB stars. The color points represent the star velocities and the gray the HI. The lines represent the median rotation velocity for a 0.5~kpc bin. The solid line is for the stars and the dashed the gas. Each panel gives the median AD. There is a significant amount of scatter and substructure in both the left and right panels, and both show similar trends of increasing AD with stellar age.\label{fig:HI_rcs}}
\end{figure}

\begin{figure}[h!]
\epsscale{1.1}
\plotone{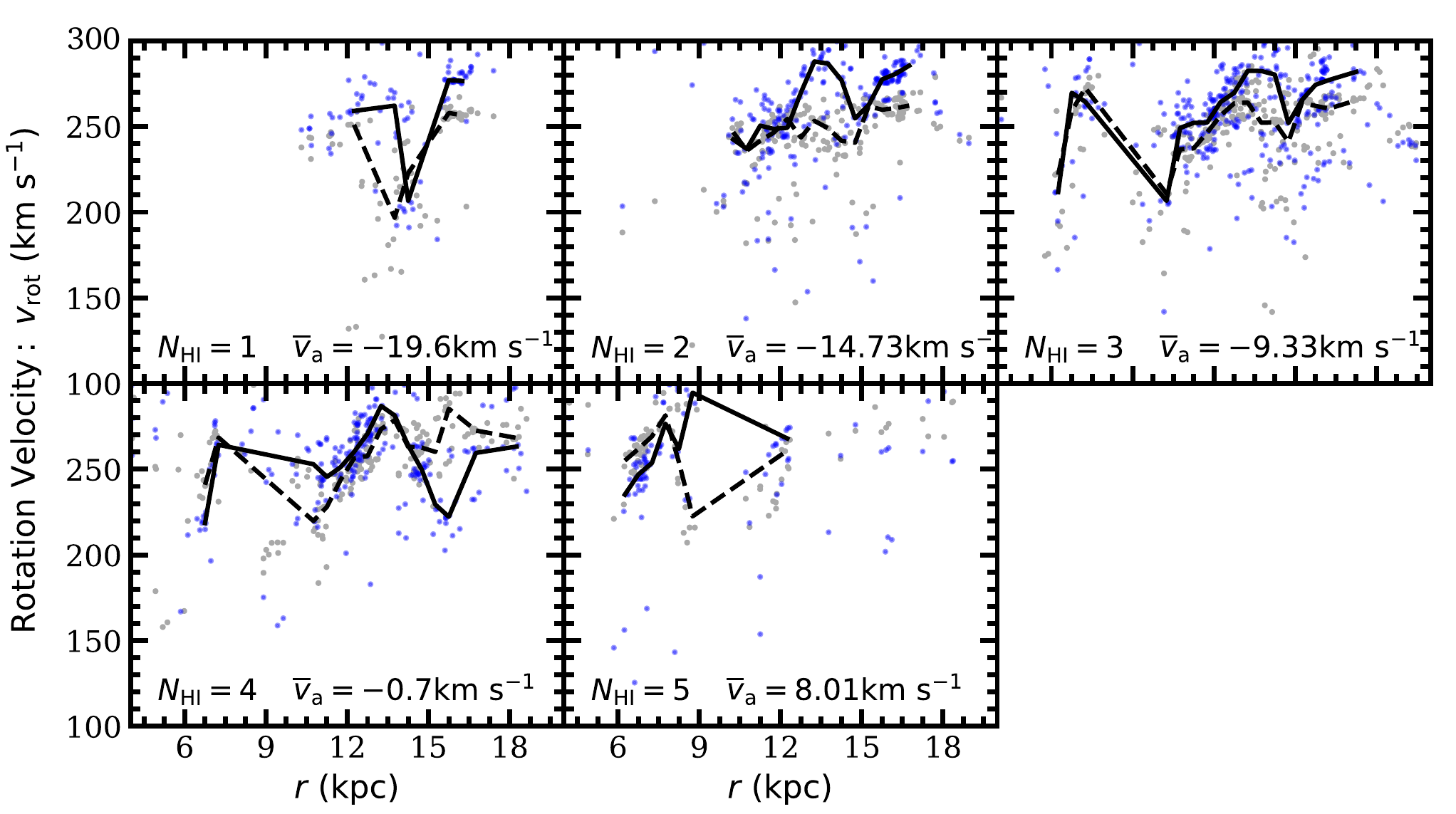}
\caption{Same as Figure~\ref{fig:RG_rcs}, except the blue points represent massive MS star velocities and the gray points the HI. The median lines are only drawn over regions that contain at least five stars per 0.5~kpc bin. \label{fig:MS_rcs}}
\end{figure}

\begin{figure}[h!]
\epsscale{1.1}
\plotone{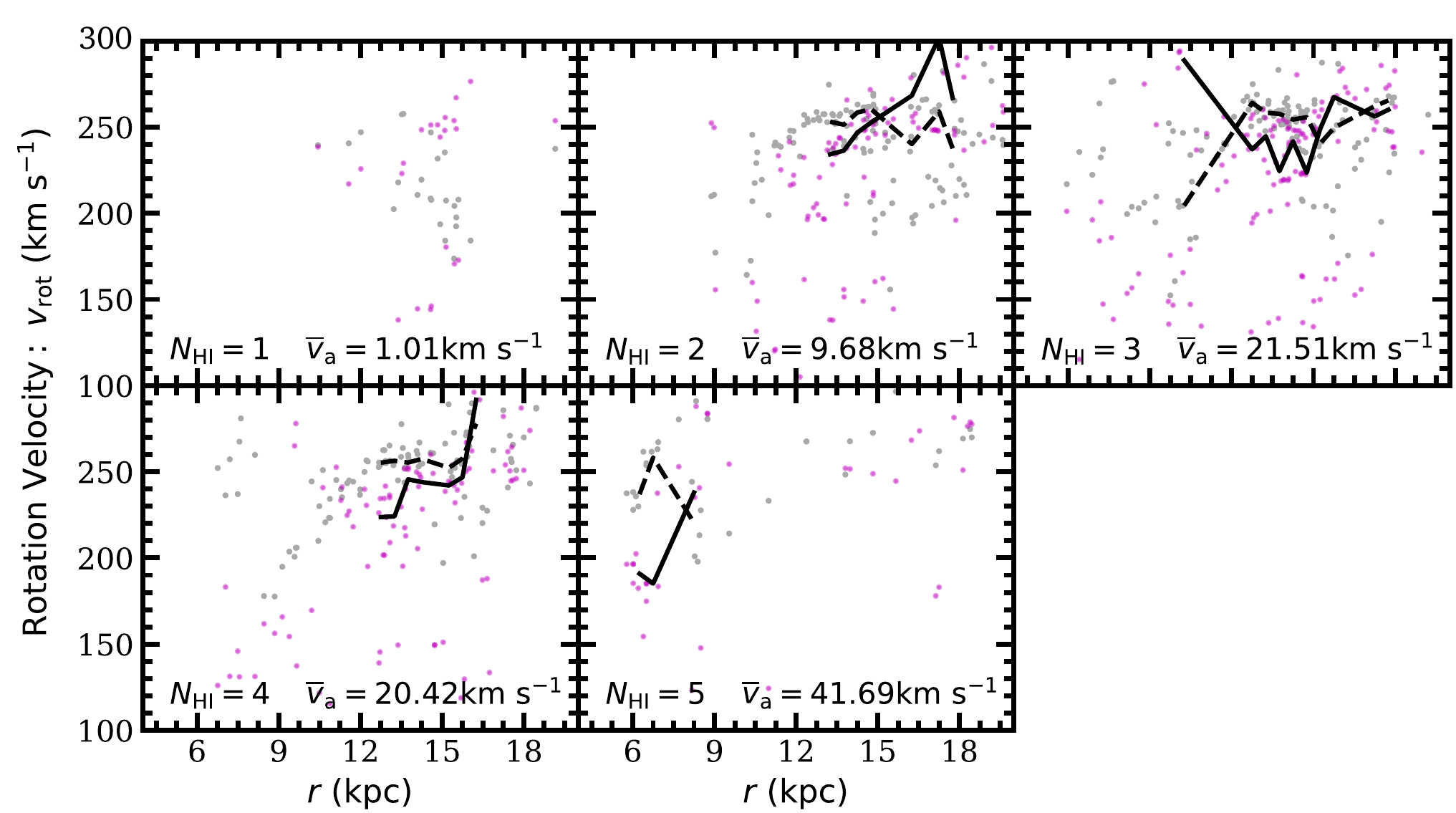}
\caption{Same as Figure~\ref{fig:RG_rcs}, except the purple points represent intermediate mass young AGB star velocities and the gray points the HI. The median lines are only drawn over regions that contain at least five stars per 0.5~kpc bin.}
\end{figure}

\begin{figure}[h!]
\epsscale{1.1}
\plotone{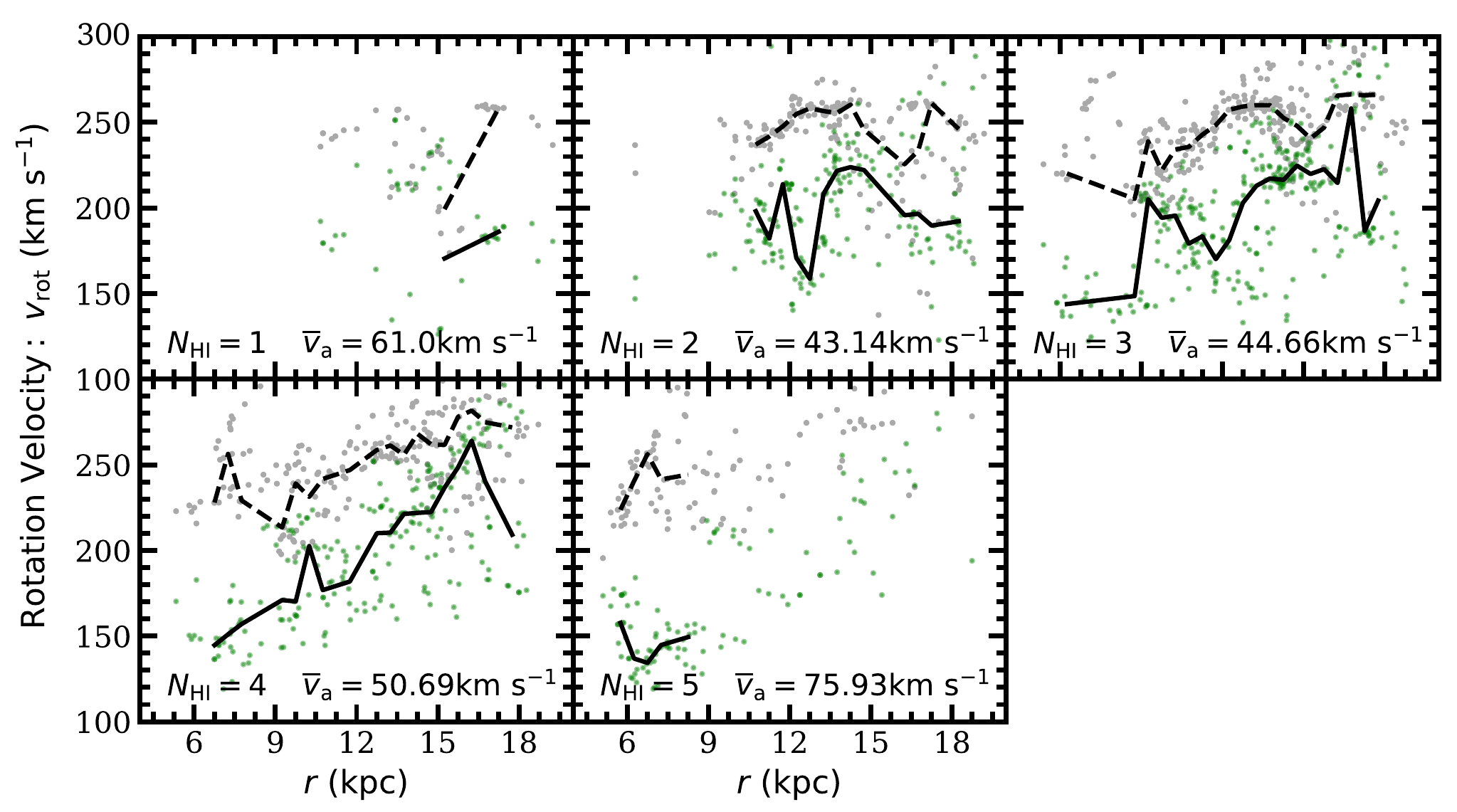}
\caption{Same as Figure~\ref{fig:RG_rcs}, except the green points represent intermediate mass older AGB star velocities and the gray points the HI. The median lines are only drawn over regions that contain at least five stars per 0.5~kpc bin.}
\end{figure}

\newpage
\subsection{Effects of Dusty Sightlines} \label{sec:dust}
\par Next we examine the effects of dust on the substructure in the RCs. Dust extinction biases us to observe more stars on the near side of the dust disk. Because of warps in M31 and patchiness in the dust layer, the radial location of the near side stars changes drastically as you move across the disk. 
\par
\begin{figure}[h!]
\epsscale{1.}
\plotone{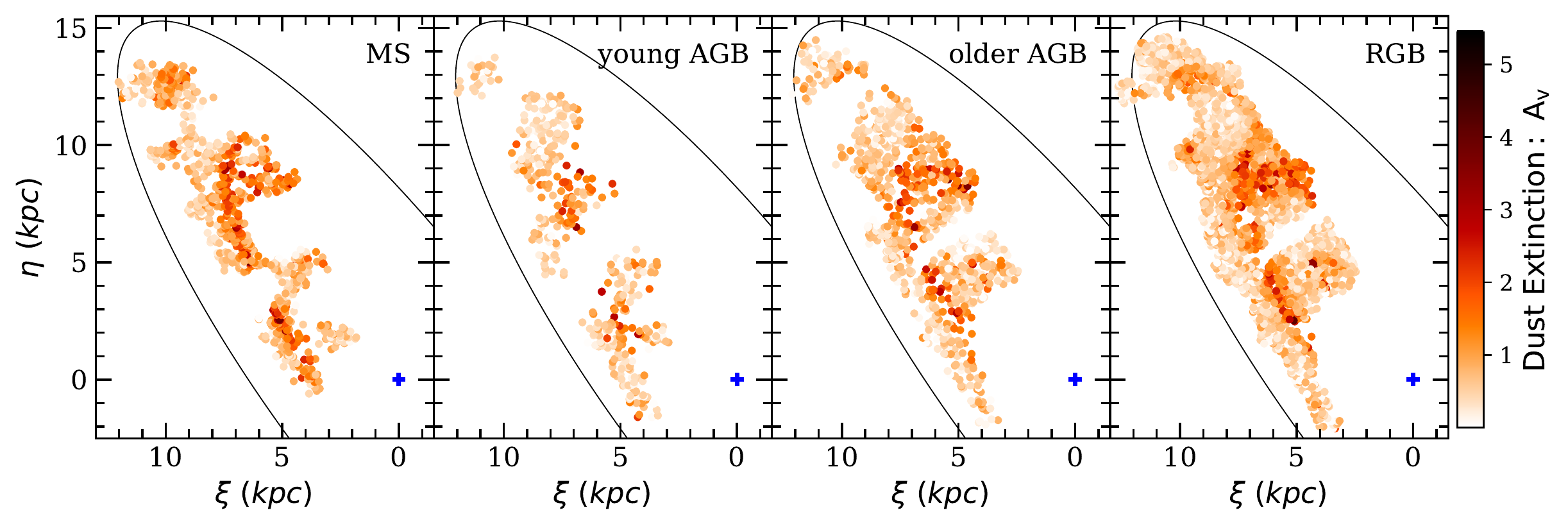}
\caption{Dust extinction maps for the four age bins. Left to right: massive MS stars, young AGB stars, older AGB stars, and RGB stars. Color represents extinction values as measured in \cite{Dalcanton2015}. The center of M31 is marked with the blue plus symbol in each panel. The ellipse is for visual reference. \label{fig:dust_maps}}
\end{figure}

\begin{figure}[h!]
\epsscale{0.7}
\plotone{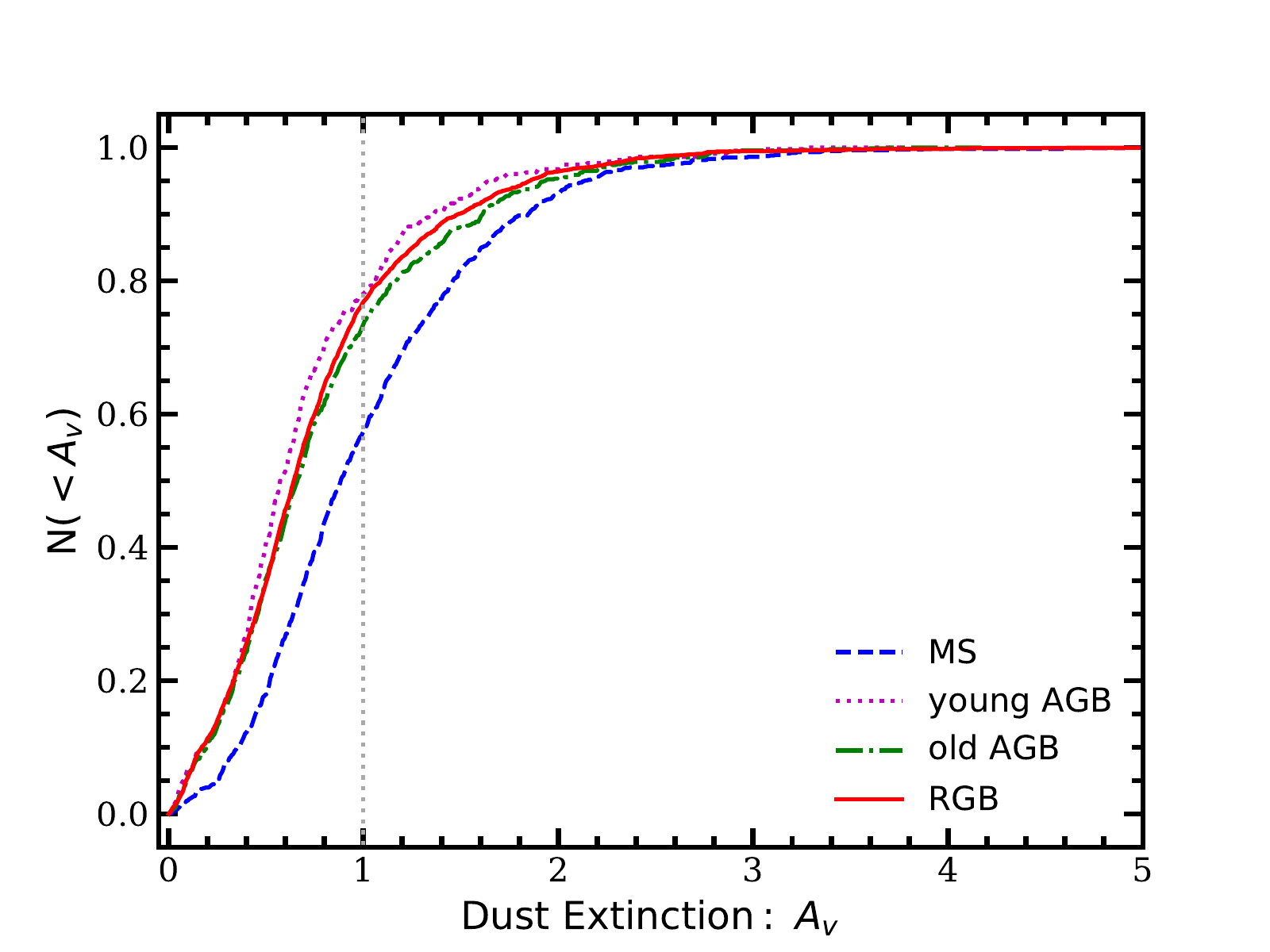}
\caption{Cumulative extinction values for each age bin. The MS stars are represented by the dashed blue line, the young AGB stars by the dotted purple line, the older AGB stars by the dashed dotted green line, and the RGB stars by the solid red line. An extinction value of $A_{v}~=~1$ is denoted by the dotted line. This will be referenced in Figure~\ref{fig:dusty_rcs}. The massive MS stars have the highest fraction of stars with $A_{v}~>~1$. \label{fig:dust_hist}}
\end{figure}

\begin{figure}[h!]
\epsscale{1.35}
\plotone{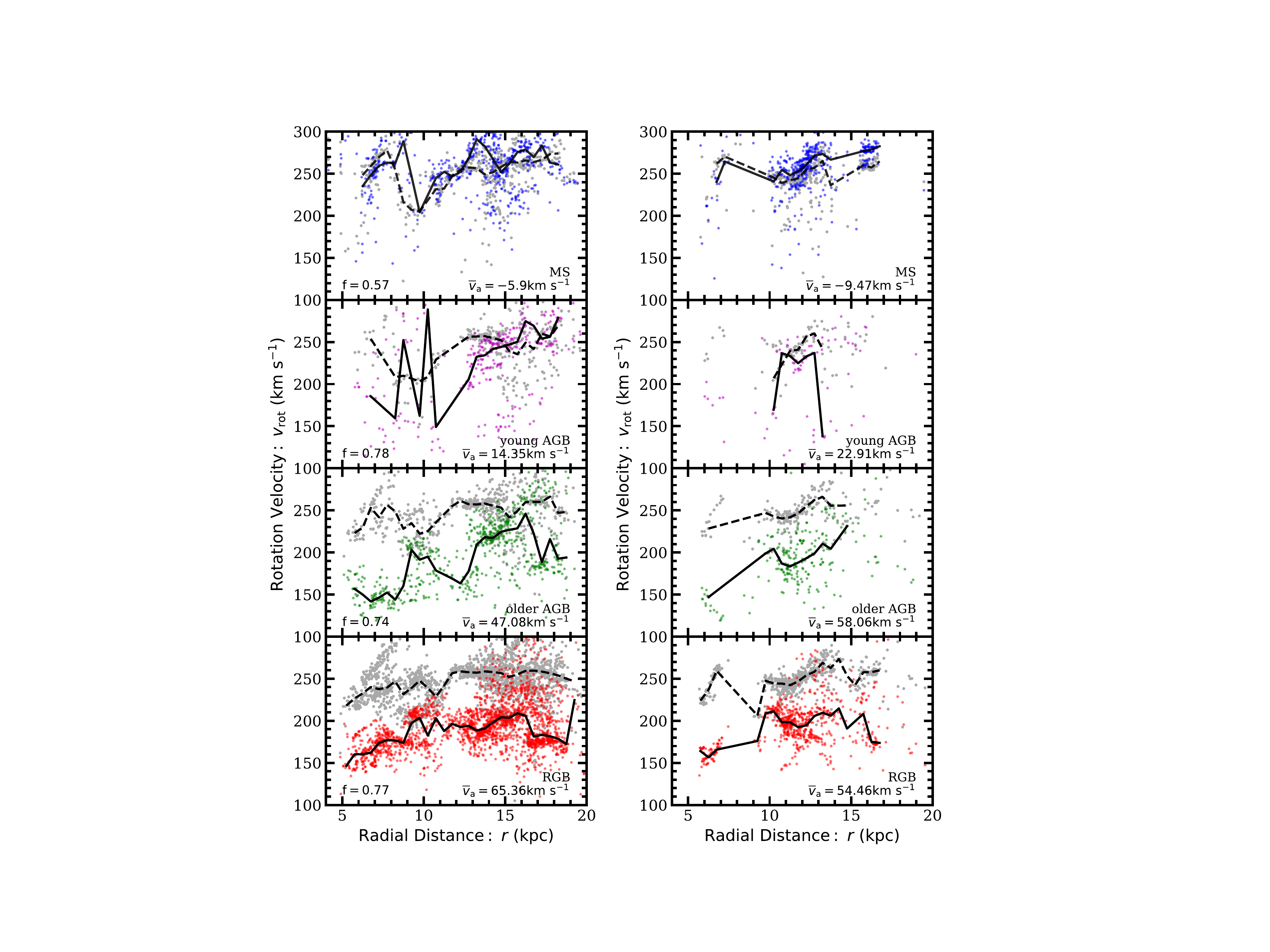}
\caption{Left panel: the RCs for sightlines where the $A_{v}~<~1$. The fraction of total stars for the age bin with $A_{v}<$ 1 are written in the left panels. The right panel is the RCs for sightlines where $A_{v}~\geq~1$. From top to bottom: massive MS stars, young AGB stars, older AGB stars, and RGB stars. In each panel, the gray points represent the HI and the color points represent the stars. The lines represent the median rotation velocity for a 0.5~kpc bin. The solid line represents the stars, and the dashed represents the gas. The median lines are only drawn over regions that contain at least five stars per bin. Each panel gives the median AD. Both sets of RCs show the trend of increasing AD as a function of stellar age. \label{fig:dusty_rcs}}
\end{figure}

 Figure~\ref{fig:dust_maps} shows extinction maps for the sightlines in our study, as measured by \cite{Dalcanton2015}. The median extinction value across the disk is $A_{v}~=~1.6$. Figure~\ref{fig:dust_hist} shows the cumulative fraction of stars at each extinction value. The vertical line denotes an extinction value of 1, which is later used in Figure~\ref{fig:dusty_rcs} to divide the population into clear and dusty sightlines. A larger of fraction of MS stars have a higher extinction than the other three age bins, which is consistent with the idea that stars form in dusty regions.

\par
\begin{figure}[hbp!] 
\epsscale{.9}
\plotone{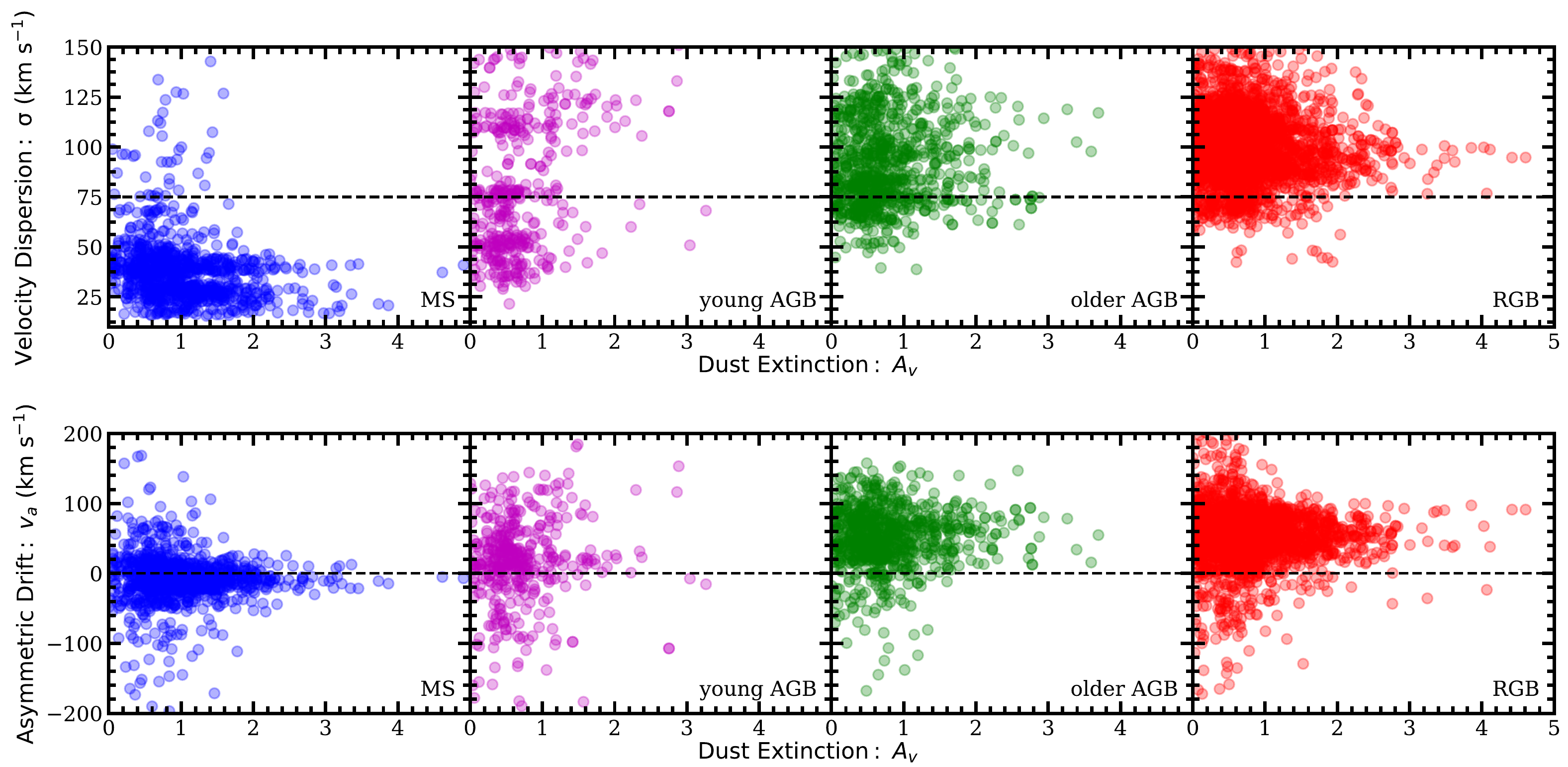}
\caption{Top panel: the velocity dispersion as a function as extinction;. The bottom panel is AD as a function of extinction. Left to right: massive MS stars, young AGB stars, older AGB stars, and RGB stars. The horizontal dotted line marks $\rm \sigma=75~km~s^{-1}$ and $\rm v_{a}~=~0~km~s^{-1}$ for reference. There do not appear to be trends between extinction and velocity dispersion (or AD) at low extinction, and there are too few stars at high extinction make conclusions about correlations. \label{fig:dust_lag_dispersion}}
\end{figure}

 Figure~\ref{fig:dust_lag_dispersion} shows dust extinction against velocity dispersion and AD. Lower velocity dispersion is expected in areas of high extinction because in these areas, a smaller number of stars in front of the dust disk are sampled instead of probing stars both in front and behind the dust layer. However, as seen in Figure~\ref{fig:dust_lag_dispersion}, there are too few stars at high extinction to see if there are correlations between the dust extinction, dispersion, and lag. There appear to be no trends amongst these values at lower extinction, which suggests this study is not significantly affected by dust extinction. Figure~\ref{fig:dusty_rcs} shows RCs for clear ($A_{v}~<~1$) and dusty ($A_{v}~\geq~1$) sightlines. Most of the sightlines, correspond to the stars in our sample have $A_{v}<1$. There appears to be no systematic difference in the AD between $A_{v}<1$ and $A_{v}>1$ sightlines for a given stellar subsample. There is a hint that the RC scatter is slightly smaller for the $A_{v}>1$ sightlines. While it may be tempting to interpret this lower RC scatter as the result of viewing only the near side of the stellar disk, we point out two caveats: (1) the young MS stars have a low AD and are presumably in a thin disk but nevertheless display this lower RC scatter for $A_{v}>1$ sightlines, and (2) the HI, which is unaffected by dust extinction, also shows lower RC scatter for $A_{v}>1$ sightlines. Instead, it is possible that the reddened sightlines happen to correspond to parts of the stellar and gas disk that have slightly more ordered kinematics than average. However, the analysis in this section is limited to examining dusty and clear sightlines and is unable to test if an individual star is behind or in front of the dust. 

\newpage
\clearpage
\subsection{Contamination from Reddened Young Hot Stars} \label{sec:RGB_red}
\par In the analysis done in the previous section, we are limited to clear or dusty sightlines instead of examining the reddening of individual stars, which would be ideal. Knowing if a star has been reddened can clarify potential discrepancies: for example, young hot stars that are behind dust will appear redder and as a result can be misclassified as older stars, and thus the higher velocity dispersion seen in the RGB bin in Figure~\ref{fig:maps} could be an artifact of a small population of younger stars in the RGB population.  
\par
\begin{figure}[h!]
\epsscale{.4}
\plotone{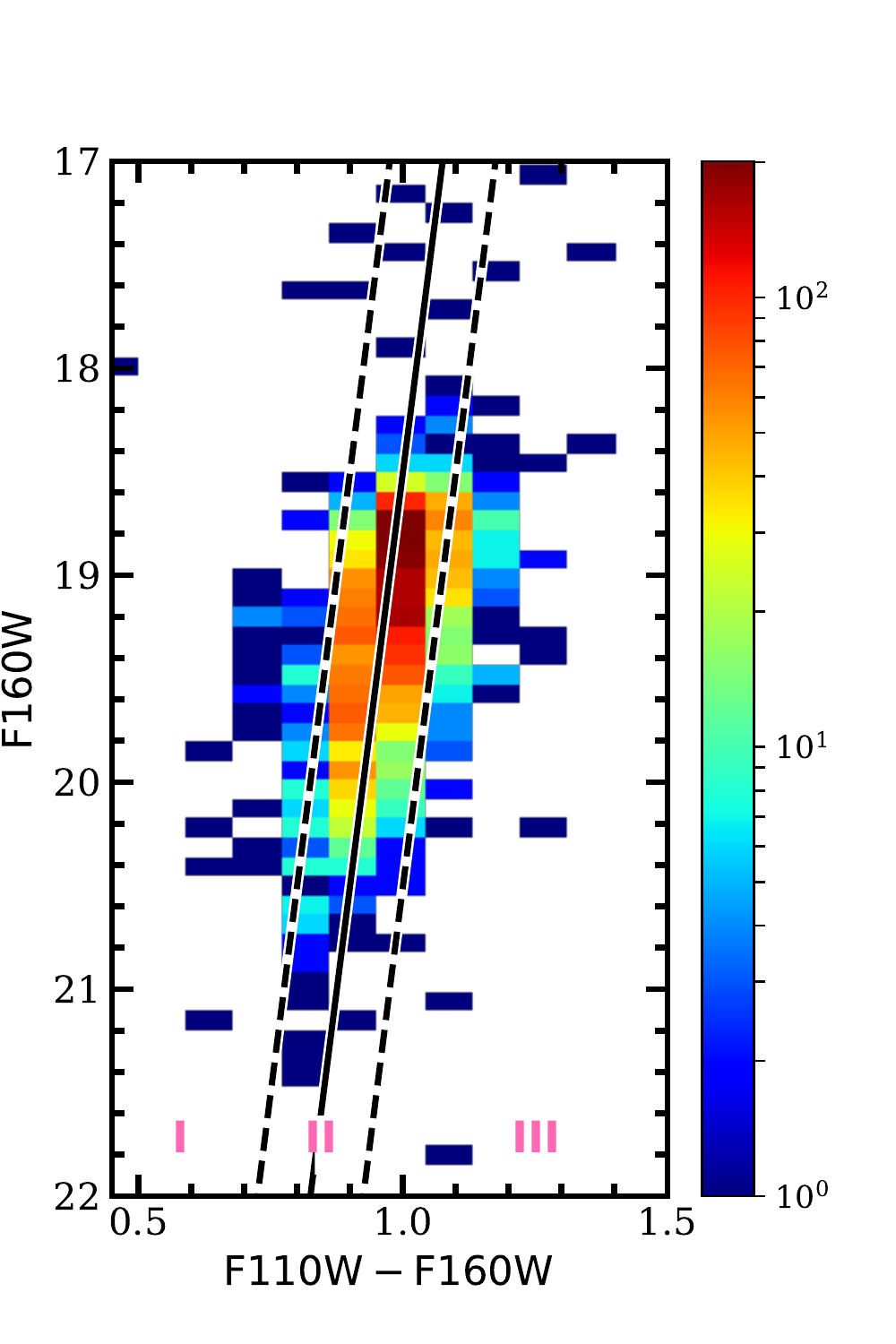}
\caption{The CMD for the RGB stars using the F110W and F160W HST filters (corresponding to \it J \rm and \it H \rm). The solid line represents unreddened stars \citep{Dalcanton2015}. Most of the RGB stars sampled are well described by the unreddened line (Region~II). Region~III contains stars that are least likely to be reddened hot young stars. Points to the far left of the solid line (Region~I) could be younger stars that have experienced reddening and were mistakingly classified as RGB stars. \label{fig:RG_CMD}}
\end{figure}

 While we cannot know the reddening of all of the stars in this study, the RGB population can be more closely examined. \cite{Dalcanton2015} analyze a near-infrared CMD of RGB stars in M31 to look for evidence that an individual star has been reddened. We use the same unreddened line from their study to determine if RGB stars in our sample have experienced reddening and therefore could be a mislabeled younger star. Figure~\ref{fig:RG_CMD} shows the CMD for the RGB star population using the F110W and F160W HST filters (corresponding to \it J \rm and \it H \rm). Most stars lie on the solid line, which denotes unreddened stars. We split the CMD into three regions, as shown in Figure~\ref{fig:RG_CMD}. Region~I is most likely to contain stars that have been reddened and therefore could be misclassified young hot stars, and Region~III is least likely to have reddened stars. 

\par
\begin{figure}[htp!]
\epsscale{.85}
\plotone{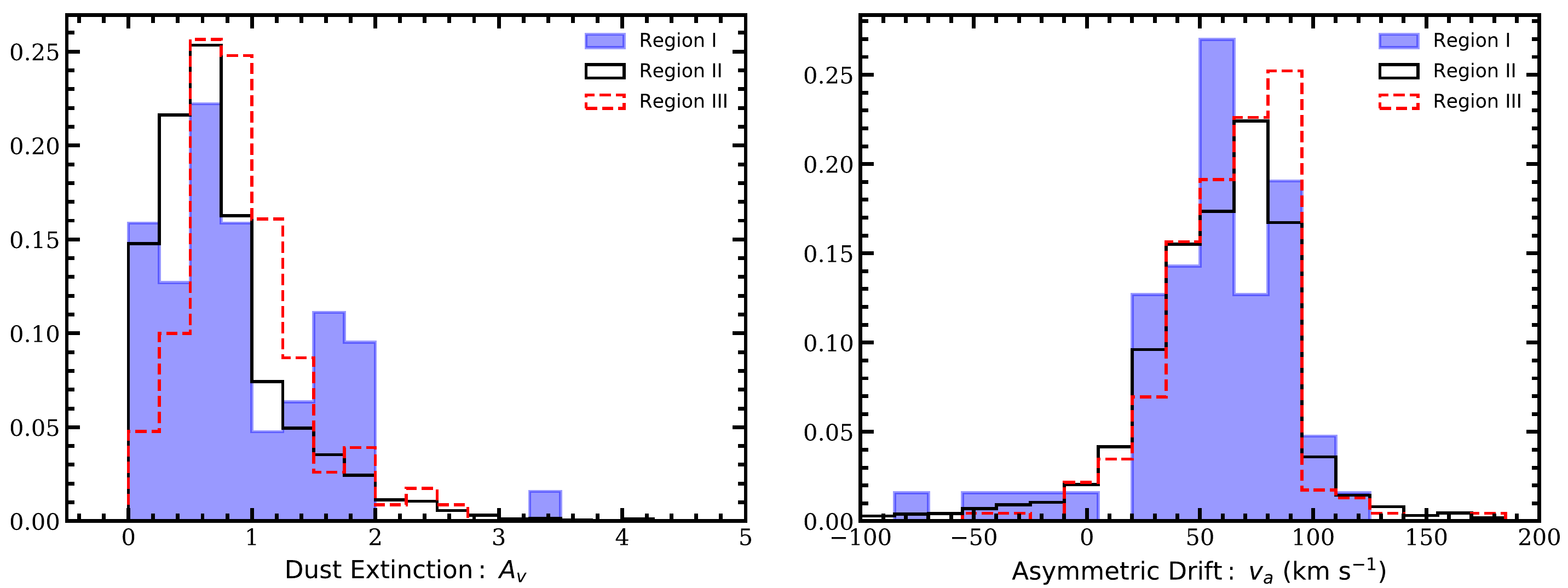}
\caption{Histograms of characteristics for the three RGB CMD regions as denoted in Figure~\ref{fig:RG_CMD}. Left panel: the line of sight dust extinction values. Right panel: AD. Region~I is filled blue histogram, Region~II is solid line black histogram, and Region~III is the dashed line red histogram. Region~I has a secondary peak at high extinction and a smaller magnitude of AD than the other two regions, which suggests it could have reddened young stars.\label{fig:RGB_hist}}
\end{figure}

 Figure~\ref{fig:RGB_hist} shows characteristics of the the three regions, as denoted in Figure~\ref{fig:RG_CMD}. The left panel of Figure~\ref{fig:RGB_hist} shows the extinction values in each of the regions. Region~I has a bimodal line of sight extinction distribution: the secondary peak is at a higher extinction than that of the peaks of the other two regions. Younger reddened stars could be occupying this peak. The right panel of Figure~\ref{fig:RGB_hist} shows the distribution of AD for each region. The AD for Region~I peaks at a lower value than for the other two regions. The AD for Region~I is significantly greater in magnitude than the AD for the massive MS population, but the fact that it is lower than the median AD for Region~II and Region~III mimics the median lag value for the MS population more than that Region~II and Region~III do. This suggests there could be contamination from younger stars in the sample we have classified as RGB stars. 

\begin{figure}[h!]
\epsscale{.5}
\plotone{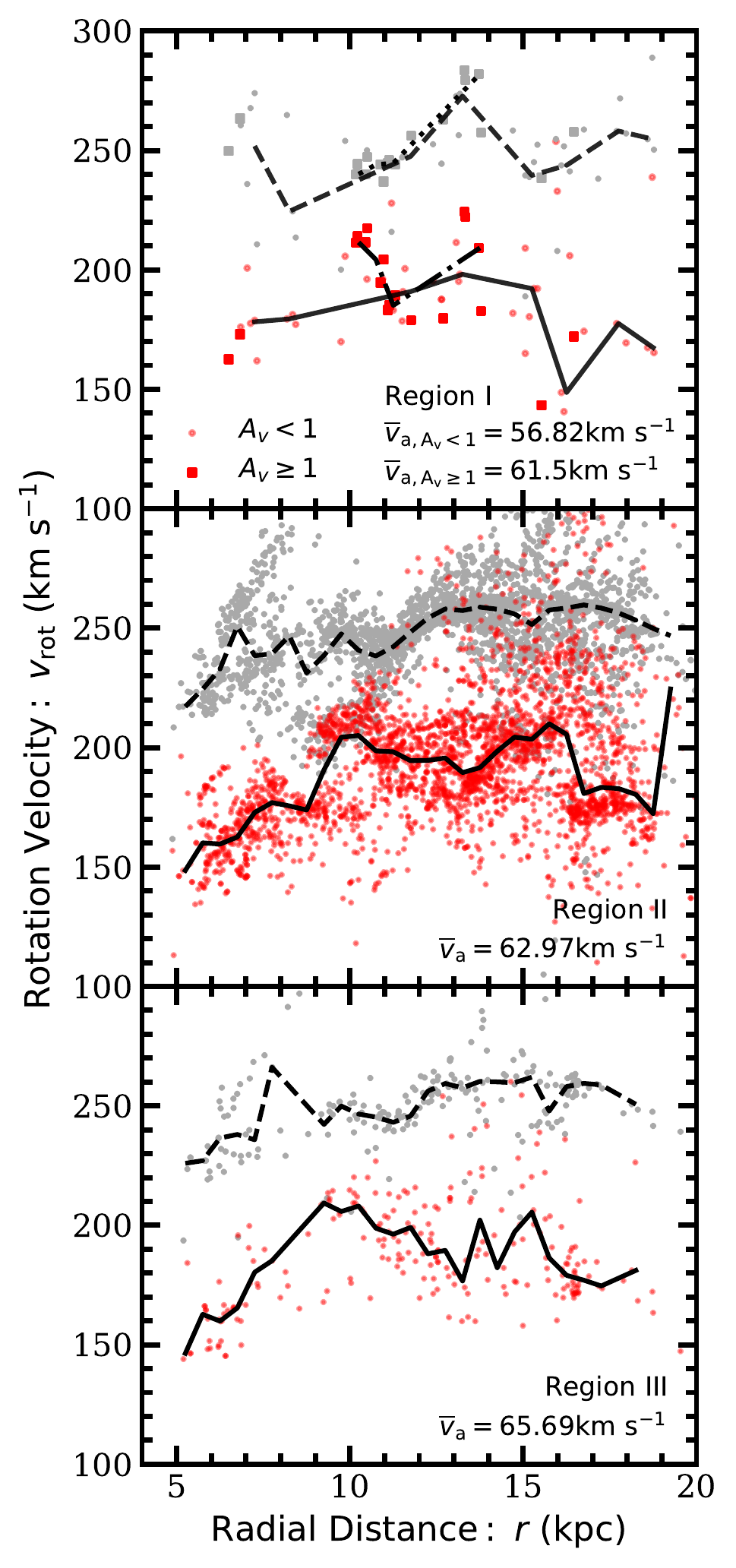}
\caption{RCs for the RGB stars in the three RGB CMD regions as denoted in Figure~\ref{fig:RG_CMD}. From top panel to bottom: Region~I, Region~II, and Region~III. In each panel, the red points represent the star velocities, and the gray represent the HI. The lines represent the median rotation velocity for a 0.5~kpc bin. In the middle and bottom panels, the solid line is for the stars and the dashed the gas. The median lines are only drawn over regions that contain at least two stars for bin for the top panel and five stars per bin for the second and third panel. In the top panel, the solid line represents the stars in Region~1 with $A_{v}~<~1$, and the dotted dashed line represents the stars in Region~1 with $A_{v}~\geq~1$. The dashed line represents the gas in this region along clear lines of sights, and the dotted line is for the gas along dusty lines of sight. The large square markers in the top panel represent stars in Region~I that have $A_{v}~\geq~1$, as seen in the left panel of Figure~\ref{fig:RGB_hist}, and could be misclassified young hot stars. However, they have an AD value that is consistent with the rest of the RGB population. \label{fig:RGB_regions_rc}}
\end{figure}

The RCs for each of the three regions are shown in Figure~\ref{fig:RGB_regions_rc}. Stars in Region~I with $A_{v}~\geq~1$ are marked with a larger square symbol in the top panel. These stars lie in the secondary peak for Region~I in the left panel of Figure~\ref{fig:RGB_hist} that corresponds to a greater extinction value. The rotational velocities of these stars are not significantly different from the velocities of stars in Region~I with clear line of sight, Region~II, and Region~III. They are consistent with the RGB AD: within 2\% of the median AD value found for the entire population. Furthermore, they are inconsistent with the behavior of the MS population, as they still lag behind the gas. Thus, while Figure~\ref{fig:RGB_hist} suggests there could be young hot stars misclassified as older stars, the RCs suggest these stars are anomalies but not younger stars. There are also so few stars in Region~I that are potentially reddened hot young stars compared to the total RGB population (0.39\% of the total RGB population) that even if they were young reddened stars, their contamination is insignificant and not a cause of the substructure in the RGB RC.

\end{document}